\documentclass[preprint,amsmath,amssymb,aps,eqsecnum,nofootinbib]{revtex4}
\usepackage{bm}
\usepackage{mathrsfs}
\usepackage[dvipdfmx]{graphicx}
\usepackage{color}
\usepackage{ulem}

\begin{document}
\title{The Aretakis constants and instability in general spherically symmetric extremal black hole spacetimes:~higher multipole modes,  late-time tails, and geometrical meanings
}
\author{Takuya Katagiri}
\author{Masashi Kimura}
\affiliation{Department of Physics, Rikkyo University, Toshima, Tokyo 171-8501, Japan}
\date{\today}
\preprint{RUP-21-19}

\begin{abstract}
We study late-time behaviors of massive scalar fields in general static and spherically symmetric extremal black hole spacetimes in arbitrary dimensions. We show the existence of conserved quantities on the extremal black hole horizons for specific mass squared and multipole modes of the scalar fields. Those quantities on the horizon are called the Aretakis constants and are constructed from the higher-order derivatives of the fields. Focusing on the region near the horizon at late times, where is well approximated by the near-horizon geometry, we show that the leading behaviors of the fields are described by power-law tails. The late-time power-law tails lead to the Atetakis instability: blowups of the transverse derivatives of the fields on the horizon.  
We further argue that the Aretakis constants and instability correspond to respectively constants and blowups of components of covariant derivatives of the fields at the late time in the parallelly propagated null geodesic frame along  the horizon. We finally discuss the relation between the Aretakis constants and ladder operators constructed from the approximate spacetime conformal symmetry near the extremal black hole horizons.

\end{abstract}

\maketitle

\section{Introduction}
Extremal black holes have long played an important role in various aspects. They have zero Hawking temperature and are expected to bring us valuable insights into the black hole thermodynamics~\cite{Bekb,Bekenstein,Hawking,Strominger:1996sh,Charmousis:2010zz,Dolan:2011xt} and the Hawking radiation~\cite{Hawking:1974sw,Angheben:2005rm}. In the context of astrophysics, it is suggested that many astrophysical black holes are nearly extremal~\cite{Volonteri:2004cf,McClintock:2006xd,Brenneman:2006hw,Gou:2013dna,McClintock:2013vwa}, and high energy phenomena around such black holes are discussed, e.g., in~\cite{Komissarov:2007rc, Banados:2009pr,Harada:2014vka}. For understanding the nature of the extremal black holes, it is important to investigate the dynamical properties of test fields and particles around them. 

Aretakis~\cite{Aretakis:2011ha,Aretakis:2011hc} has discussed late-time behaviors of test massless scalar fields in four-dimensional extremal Reissner-Nordstr\"{o}m black holes. When generic initial data are given on an initial hypersurface crossing the horizon, he argued that the higher-order transverse derivatives of the fields blow up polynomially in time, not exponential, on the event horizon, while they decay outside the horizon. This blowup on the horizon is called the Aretakis instability. The occurrence of the Aretakis instability is associated with the fact that late-time behaviors of fields are described by power-law tails~\cite{Lucietti:2012xr,Ori:2013iua,Sela:2015vua,Bhattacharjee:2018pqb,Angelopoulos:2018uwb,Price:1971fb}, not the exponential decay in time. The instability also occurs against vector, tensor, and massive or charged scalar fields~\cite{Lucietti:2012xr,Lucietti:2012sf,Zimmerman:2016qtn}, and in other spacetimes such as extremal Kerr(-Newman)~\cite{Lucietti:2012sf,Aretakis:2012ei}, extremal Ba\~{n}ados-Teitelboim-Zanelli~\cite{Gralla:2019isj}, and two-dimensional anti-de Sitter spacetimes (${\rm AdS}_2$)~\cite{Lucietti:2012xr,Hadar:2017ven,Katagiri:2021scx}\footnote{It has been argued that in~\cite{Lucietti:2012xr} the Aretakis instability in global ${\rm AdS}_2$ is just a coordinate effect, while in~\cite{Hadar:2017ven} the Aretakis instability in the near-horizon geometry of extremal black holes is not the coordinate effect. In~\cite{Katagiri:2021scx}, contrary to the claim in~\cite{Lucietti:2012xr}, it has been shown that the late-time divergent behavior in global ${\rm AdS}_2$ has a geometrical meaning: blowups of some component of covariant derivatives of fields in the parallelly propagated null geodesic frame.}, and in higher dimensions~\cite{Murata:2012ct}. The nonlinear evolution of the Aretakis instability has been investigated in~\cite{Murata:2013daa,Hadar:2018izi,Angelopoulos:2019gjn}.

Related to the Aretakis instability, conserved quantities along the horizons, which are constructed from the higher-order radial derivatives of the fields, are studied~\cite{Aretakis:2011ha,Aretakis:2011hc}.
These conserved quantities on the horizon are called the Aretakis constants.
When the Aretakis constants contain $(N+1)$th- and lower order derivatives, $(N+2)$th- and higher-order derivatives on the horizon blow up at late times, which is the Aretakis instability. The existence of the Aretakis constants has been argued in various setups~\cite{Aretakis:2011hc,Lucietti:2012xr,Lucietti:2012sf,Murata:2012ct}. In~\cite{Lucietti:2012xr,Bizon:2012we,Godazgar:2017igz,Bhattacharjee:2018pqb,Cvetic:2018gss}, it has been shown that there exists one-to-one correspondence of the Aretakis constants and the Newman-Penrose constants~\cite{Newman:1968uj} in four-dimensional extremal Reissner-Nordstr\"{o}m black holes. The relation with asymptotic symmetries near degenerate Killing horizons is suggested in~\cite{Akhmedov:2017ftb,Godazgar:2018vmm}. For the ${\rm AdS}_2$ case, the relation with the spacetime conformal symmetry in ${\rm AdS}_2$ is pointed out~\cite{Cardoso:2017qmj,Katagiri:2021scx}. 

This paper aims to obtain a deeper understanding of the properties of the Aretakis constants and instability. We consider the massive Klein-Gordon field in general static and spherically symmetric extremal black holes in arbitrary dimensions. We investigate the existence of conserved quantities, i.e., the Aretakis constants. We further study late-time behaviors of the field near the black hole horizon, namely, the Aretakis instability, in terms of the near-horizon geometry~\cite{Kunduri:2013gce}. We also discuss the geometrical properties of the Aretakis constants and instability: constants and blowups of components of covariant derivatives of the fields at late times in the parallelly propagated null geodesic frame along the horizons. We finally study the construction of the Aretakis constants in terms of approximate ${\rm AdS}_2$ symmetry near the extremal black hole horizons.

The behavior of test fields in the near-horizon geometry can be reduced to scalar fields on ${\rm AdS}_2$. In the pure ${\rm AdS}_2$ case, it has been shown that the Aretakis constants of the massive Klein-Gordon field can be derived from ladder operators associated with the spacetime conformal symmetry~\cite{Cardoso:2017qmj,Katagiri:2021scx}. We expect that the Aretakis constants in the extremal black holes can be derived from the ladder operators associated with the approximate ${\rm AdS}_2$ symmetry near the extremal black hole horizons.\footnote{\label{footnote:intro2}In Appendix~F in~\cite{Cardoso:2017qmj}, the construction of the Aretakis constants of the extremal Reissner-Nordstr\"om black holes from the ${\rm AdS}_2$ symmetry was discussed. However, the conserved quantities on the black hole horizon for $\ell \ge 1$ discussed in~\cite{Cardoso:2017qmj} vanish, thus they are not the Aretakis constants for $\ell \ge 1$. In the present paper, we provide more careful analysis and derive the Aretakis constants from the approximate ${\rm AdS}_2$ symmetry.}

This paper is organized as follows. In Sec.~\ref{sec:AretakisconstinEBH}, we briefly review the Aretakis constants and instability for the massless Klein-Gordon field in four-dimensional extremal Reissner-Nordstr\"{o}m black holes based on~\cite{Aretakis:2011ha,Aretakis:2011hc,Lucietti:2012xr}. After that, we investigate whether there exist the same kind of quantities for massive Klein-Gordon fields in general static and spherically symmetric extremal black holes in arbitrary dimensions. In Sec.~\ref{sec:latetimepowerlawtailsandinstability}, we investigate late-time behaviors of the field in the extremal black hole spacetimes. In Sec.~\ref{sec:AretakisconstandinstabilityinPP}, we explain the Aretakis constants and instability in the parallelly propagated null geodesic frame along the horizon. In Sec.~\ref{sec:Aretakisconstsfromladder}, we discuss the Aretakis constants in terms of ladder operators constructed from approximate ${\rm AdS}_2$ symmetries near the extremal black hole horizons. In the final section, we summarize this paper. Appendices give detailed calculations and explicit examples for the main text. 

\section{Klein-Gordon fields and the Aretakis constants in extremal black holes}
\label{sec:AretakisconstinEBH}

\subsection{Aretakis constants of massless Klein-Gordon fields in four-dimensional extremal Reissner-Nordstr\"om black holes}
\label{subsec:AretakisconstinEBH-1}
We first briefly review conserved quantities on the event horizon of four-dimensional extremal Reissner-Nordstr\"om spacetimes based on~\cite{Aretakis:2011ha,Aretakis:2011hc,Lucietti:2012xr}. In the ingoing Eddington-Finkelstein coordinates $\left(v,r,\theta,\varphi\right)$, the four-dimensional extremal Reissner-Nordstr\"om spacetime is described by
\begin{equation}
\label{OERN}
ds^2=-\left(1-\frac{r_H}{r}\right)^2dv^2+2dvdr+r^2d\Omega^2,
\end{equation}
where $d\Omega^2=d\theta^2+\sin^2\theta d\varphi^2$. The event horizon is located at $r=r_H$. For later convenience, we perform a coordinate transformation
\begin{equation}
\rho=r-r_H.
\end{equation}
Then, the line element~\eqref{OERN} is rewritten as
\begin{equation}
\label{ERN}
ds^2=-\frac{\rho^2}{\left(r_H+\rho\right)^2}dv^2+2dvd\rho+\left(r_H+\rho\right)^2d\Omega^2.
\end{equation}
By definition of $\rho$, the event horizon is located at $\rho=0$, and the exterior region corresponds to $\rho>0$. 

On the extremal Reissner-Nordstr\"om spacetime, we consider the massless Klein-Gordon field $\Phi(v,\rho,\theta,\varphi)$ obeying
\begin{equation}
\label{KGE}
\nabla_\mu\nabla^\mu\Phi=0.
\end{equation}
With the spherical symmetry of the spacetime, expanding the field in the scalar harmonics~$\mathbb{S}_{\ell}\left(\theta,\varphi\right)$ as 
\begin{equation}
\label{anz}
\Phi=\phi_{\ell}\left(v,\rho\right)\mathbb{S}_{\ell}\left(\theta,\varphi\right),
\end{equation}
where $\mathbb{S}_\ell$ is a regular solution of 
\begin{equation}
\left[\hat{\Delta}+\ell(\ell+1)\right]\mathbb{S}_{\ell}=0,
\end{equation}
with $\ell=0,1,2,\cdots$, and $\hat{\Delta}$ is the Laplace operator on the two-dimensional unit sphere, we obtain an equation for $\phi_{\ell}$,
\begin{equation}
\label{eom}
\partial_\rho\left(\rho^2\partial_\rho\phi_{\ell }\right)+2\left(r_H+\rho\right)\partial_v\partial_\rho\left[\left(r_H+\rho\right)\phi_{\ell }\right]-\ell(\ell+1)\phi_{\ell}=0.
\end{equation}
Acting the $\ell$th-derivative operator with respect to $\rho$, $\partial_{\rho}^{\ell}$, on Eq.~\eqref{eom} and evaluating it at the event horizon $\rho=0$, we obtain
\begin{equation}
\label{con1}
\left.\partial_v\partial_{\rho}^{\ell}\left[\left(r_H+\rho\right)\partial_\rho\left\{\left(r_H+\rho\right)\phi_\ell\right\}\right]\right|_{\rho=0}=0.
\end{equation}
We see that quantities defined by
\begin{equation}
\label{arec}
\left.\mathcal{H}^{RN}_\ell:=\partial_{\rho}^{\ell}\left[\left(1+\frac{\rho}{r_H}\right)\partial_\rho\left\{\left(1+\frac{\rho}{r_H}\right)\phi_\ell\right\}\right]\right|_{\rho=0},
\end{equation}
are independent of $v$, i.e., $\partial_v\mathcal{H}^{RN}_\ell=0$. The quantities $\mathcal{H}^{RN}_\ell$ are conserved along the event horizon, and called the \textit{Aretakis constants}~\cite{Aretakis:2011ha,Aretakis:2011hc,Lucietti:2012xr}. According to~\cite{Aretakis:2011ha,Aretakis:2011hc}, the existence of the Aretakis constants implies the polynomial growth of $\partial_\rho^{k(\ge \ell+2)}\phi_\ell|_{\rho=0}$ in $v$ at late times~$v\to\infty$. This is called the \textit{Aretakis instability}. The Aretakis instability does not directly imply blowups of physical quantities. In fact, the second-and higher-order, not the first-order, transverse derivatives of the field on the horizon diverge at the late time; therefore the energy-momentum tensor does not blow up. Note that the derivatives of the fields lower than $\ell$ decay on and outside the horizon at the late time~\cite{Aretakis:2011ha,Aretakis:2011hc,Lucietti:2012xr}.

For the generic configuration of fields, which is expressed as a superposition of multipole modes, the first-order transverse derivative of the field on the horizon approaches  the Aretakis constant for $\ell=0$ in Eq.~\eqref{arec} at late times. This implies that the radial-radial component of the energy-momentum tensor of the field on the horizons does not decay at late times. The Aretakis constant is physically interpreted as ``horizon hair"  in the sense of the energy density measured by an infalling observer at the horizon~\cite{Lucietti:2012xr}. In~\cite{Angelopoulos:2018yvt,Burko:2020wzq}, the observability of the Aretakis constants and instability have been argued.

\subsection{Aretakis constants of massive Klein-Gordon fields in general static and spherically symmetric extremal black holes in arbitrary dimensions}
\label{subsec:AretakisconstinEBH}
We shall show that there exist the Aretakis constants for the massive Klein-Gordon field in general static and spherically symmetric extremal black hole spacetime. We now introduce a line element describing an $n$-dimensional static and spherically symmetric black hole with a degenerate Killing horizon. In the Gaussian null coordinates $\left(v,\rho,\theta^A\right)$ around the Killing horizon $\rho=0$, the line element reads as~\cite{Kunduri:2013gce}
\begin{equation}
\label{EBH}
ds^2=-\rho^2\left(\lambda_0 +\delta\lambda(\rho)\right)dv^2+2dvd\rho+\left(\gamma_0+\delta\gamma(\rho)\right)d\Omega^2_{n-2},
\end{equation} 
where $\lambda_0 $, $\gamma_0$ are positive constants, $\delta\lambda(\rho)$, $\delta\gamma(\rho)$ are functions with $\delta\lambda(0)=\delta\gamma(0)=0$, and $d\Omega^2_{n-2}=\gamma_{AB}d\theta^Ad\theta^B$ is the line element of the ($n-2$)-dimensional unit sphere, $A,B=2,3,\cdots n-1$. The constant~$\gamma_0$ is the square of the areal radius at the Killing horizon.  We assume that $\delta\lambda$ and $\delta \gamma$ can be expanded as the Taylor series,
\begin{equation}
\begin{split}
\label{labellambdagamma}
\delta\lambda=&\sum_{i=1}^\infty\lambda_i \rho^i,\\
\delta\gamma=&\sum_{i=1}^\infty\gamma_i \rho^i,\\
\end{split}
\end{equation}
where $\lambda_i$ and $\gamma_i$ are constants.\footnote{In the cases of the four-dimensional extremal Reissner-Nordstr\"om-{\rm AdS} black holes, the parameters are 
\begin{equation}
\label{ERNexample}
\lambda_0=\frac{1}{r_H^2}+\frac{6}{L^2},~~\lambda_1=-\frac{2}{r_H^3}-\frac{4}{r_H L^2},~~ \gamma_0=r_H^2,~~\gamma_1=2r_H,
\end{equation}
where $L$ and ${r_H}$ are the AdS curvature radius and the horizon radius, respectively. In the $n$-dimensional extremal Reissner-Nordstr\"om black hole cases, those are
\begin{equation}
\lambda_0=\frac{(n-3)^2}{r_H^2},~~\lambda_1=-\frac{(n-2)(n-3)^2}{r_H^3},~~ \gamma_0=r_H^2,~~\gamma_1=2r_H.
\end{equation}
 }

We consider the massive Klein-Gordon field
\begin{equation}
\label{KGE2}
\left[\nabla_{\mu}\nabla^{\mu}-\mu^2\right]\Phi=0.
\end{equation}
Expanding $\Phi$ in the scalar harmonics $\mathbb{S}_\ell$ as
\begin{equation}
\Phi=\phi_{\ell}(v,\rho)\mathbb{S}_\ell(\theta^A),
\end{equation}
we obtain an equation for $\phi_{\ell}$,
\begin{equation}
\begin{split}
\label{eom2}
&\partial_\rho\left[\rho^2\left(\lambda_0 +\delta\lambda\right)\left(\gamma_0+\delta\gamma\right)\partial_\rho\phi_\ell\right]+2\left(\gamma_0+\delta\gamma\right)^{1/2}\partial_v\partial_\rho\left[\left(\gamma_0+\delta\gamma\right)^{1/2}\phi_\ell\right]\\
&-\ell(\ell+n-3)\phi_\ell-\mu^2\left(\gamma_0+\delta\gamma\right)\phi_\ell=0.
\end{split}
\end{equation}
The function~$\mathbb{S}_\ell$ is a regular solution of 
\begin{equation}
\left[\hat{\Delta}_n+\ell(\ell+n-3)\right]\mathbb{S}_\ell=0,
\end{equation}
where $\hat{\Delta}_n$ is the Laplace operator on the $(n-2)$-dimensional unit sphere. As shown in Appendix~\ref{Appendix:A}, 
there exists a function $Z_N(\rho)$ such that
\begin{equation}
\begin{split}
&\left.\partial_\rho^{N}\left[Z_N\left\{\partial_\rho\left(\rho^2\left(\lambda_0 +\delta\lambda\right)\left(\gamma_0+\delta\gamma\right)\partial_\rho\phi_\ell\right)-N(N+1)\gamma_0\lambda_0  \phi_\ell-\mu^2\delta\gamma\phi_\ell\right\}\right]\right|_{\rho=0}= 0,
\label{eq:zequation}
\end{split}
\end{equation}
where $N$ is a nonnegative integer. The explicit form of $Z_N$ depends on the function form of $\delta\lambda$ and $\delta\gamma$.  We note that if $(\lambda_0 +\delta\lambda)(\gamma_0+\delta\gamma)=1$, e.g., Reissner-Nordstr\"om spacetime cases, $Z_N=1$.
Acting the $N$th-derivative operator~$\partial_\rho^N$ on Eq.~\eqref{eom2} and evaluating it at the horizon $\rho=0$, we obtain
\begin{equation}
\begin{split}
\label{nga}
&\left.2\partial_v\partial_{\rho}^{N}\left[Z_N \left(\gamma_0+\delta\gamma\right)^{1/2}\partial_\rho\left\{\left(\gamma_0+\delta\gamma\right)^{1/2}\phi_\ell\right\}\right]\right|_{\rho=0}\\
&=\left.\gamma_0\left\{\mu^2-\lambda_0 N(N+1)+\frac{\ell(\ell+n-3)}{\gamma_0}\right\}\partial^{N}_\rho \left(Z_{N}\phi_\ell\right)\right|_{\rho=0},
\end{split}
 \end{equation}
where we have used Eq.~\eqref{eq:zequation}. We see that quantities defined by 
\begin{equation}
\label{garec}
\mathcal{H}_{N}:=\left.\partial_{\rho}^{N}\left[Z_N \left(1+\frac{\delta\gamma}{\gamma_0}\right)^{1/2}\partial_\rho\left\{\left(1+\frac{\delta\gamma}{\gamma_0}\right)^{1/2}\phi_\ell\right\}\right]\right|_{\rho=0},
\end{equation}
are independent of $v$, i.e., $\partial_v\mathcal{H}_{N}=0$, if and only if $\mu^2=\lambda_0 N(N+1)-\ell(\ell+n-3)/\gamma_0$ or equivalently the nonnegative integer~$N$ is described by
\begin{equation}
\label{N}
N=\Delta-1,
\end{equation}
where
\begin{equation}
\label{Delta}
\Delta:=\frac{1}{2}+\sqrt{\frac{\mu^2}{\lambda_0 }+\frac{\ell(\ell+n-3)}{\lambda_0 \gamma_0}+\frac{1}{4}}.
\end{equation}
The quantities~$\mathcal{H}_{N}$ are the Aretakis constants in the present system.\footnote{We comment on the relation with previous works. The case for four-dimensional, massless and $\lambda_0 \gamma_0 = 1$ was discussed in~\cite{Aretakis:2011hc}. The case in which a spherically symmetric scalar field has specific mass squared~$\mu^2r_H^2 = N(N+1)$ in the four-dimensional extremal Reissner-Nordstr\"om spacetime of the radius~$r_H$ was made in~\cite{Lucietti:2012xr}.  The case of $N = 0$ is included in the general discussion~\cite{Murata:2012ct,Lucietti:2012sf}.
} 

For the massless Klein-Gordon field in the four-dimensional extremal Reissner-Nordstr\"om spacetime, i.e., $\mu^2=0, n=4, \lambda_0\gamma_0=1$, the nonnegative integer~$N$ corresponds to $\ell$. For generic cases, $\Delta$ in Eq.~\eqref{Delta} is not necessarily an integer, and then there are no Aretakis constants. In general, for given parameters~$\ell, N, \lambda_0, \gamma_0$, if we choose $\mu^2=\lambda_0 N(N+1)-\ell(\ell+n-3)/\gamma_0$, the Aretakis constants exist for these specific parameters. 

In particular, there are two interesting cases where the Aretakis constants exit:
(i) $\mu = 0$ and $\ell = 0$ case. Then, we can define the Aretakis constants with $N = 0$ in arbitrary dimensions~$n$.
(ii) $\mu=0$ and $\lambda_0\gamma_0=(n-3)^2$. Then, $\Delta=1+\ell/(n-3)$, and the Aretakis constants with $N= \ell/(n-3)$ exist in the case, where $\ell$ is proportional to $n-3$. Note that $\lambda_0\gamma_0=(n-3)^2$ holds for $n$-dimensional extremal Reissner-Nordstr\"om spacetimes. 

\subsection{Massive Klein-Gordon fields in the near-horizon region}
\label{subsec:MKGfiedsinNHG}
For later convenience, we shall see that the Klein-Gordon equation~\eqref{KGE2} in the near-horizon region can be reduced to the problem of massive Klein-Gordon fields in the two-dimensional anti-de Sitter spacetime (${\rm AdS}_2$). For the line element~\eqref{EBH}, performing a coordinate transformation $v=\tilde{v}/\epsilon,\rho=\epsilon \tilde{\rho}$ and taking a limit of $\epsilon\to 0$, we obtain the \textit{near-horizon geometry}~\cite{Kunduri:2013gce}
\begin{equation}
\begin{split}
\label{generalm1}
ds^2\rightarrow -\lambda_0\tilde{\rho}^2d\tilde{v}^2+2d\tilde{v}d\tilde{\rho}+\gamma_0d\Omega_{n-2}^2.
\end{split}
\end{equation}
We should note that the transformation~$v=\tilde{v}/\epsilon, \rho=\epsilon\tilde{\rho}$ and taking the limit of $\epsilon\to0$ correspond to zoom up of the late-time and the vicinity of the black hole horizon in the original coordinates $(v,\rho)$. The line element~\eqref{generalm1} is invariant under the transformation $\left(\tilde{v},\tilde{\rho}\right)\to \left(\tilde{v}/\epsilon,\epsilon \tilde{\rho}\right)$. The scaling symmetry appears near the degenerate Killing horizons of the extremal black holes at the late time. In fact, $(\tilde{v}, \tilde{\rho})$ part of the line element~\eqref{generalm1} is ${\rm AdS}_2$, which is the two-dimensional maximally symmetric spacetime with negative curvature, and hence the line element~\eqref{generalm1} has the same symmetry as ${\rm AdS}_2$. In this paper, we say that the extremal black hole~\eqref{EBH} has the approximate ${\rm AdS}_2$ symmetry in the near-horizon region.

We rewrite Eq.~\eqref{eom2} as
\begin{equation}
\label{effeom}
\left[2\partial_v\partial_\rho+\partial_\rho\left(\lambda_0 \rho^2\partial_\rho\right)-\bar{\mu}^2\right]\phi_\ell=\delta[\phi_\ell],
\end{equation}
where 
\begin{equation}
\label{barmu}
\bar{\mu}^2:=\mu^2+\frac{\ell(\ell+n-3)}{\gamma_0},
\end{equation}
and
\begin{equation}
\begin{split}
\label{deltaN}
\delta\left[\phi_\ell\right]=\frac{1}{\gamma_0}\left(-\partial_v\left[2\delta\gamma\partial_\rho\phi_\ell+\left(\partial_\rho\delta\gamma\right)\phi_\ell\right]-\partial_\rho\left[\rho^2(\gamma_0\delta\lambda+\lambda_0 \delta\gamma+\delta\lambda\delta\gamma)\partial_\rho\phi_\ell\right]+\mu^2\delta\gamma\phi_\ell\right).
\end{split}
\end{equation}
We note that the first and second terms in the square bracket of the left-hand side of Eq.~\eqref{effeom} can be written by the d'Alembertian for ${\rm AdS}_2$:
\begin{equation}
\label{BoxAdS2}
2\partial_v\partial_\rho+\partial_\rho\left(\lambda_0 \rho^2\partial_\rho\right)=:\square_{{\rm AdS}_2}.
\end{equation}
Now, performing the coordinate transformation $v=\tilde{v}/\epsilon, \rho=\epsilon\tilde{\rho}$, we can write Eq.~\eqref{effeom} in the form
\begin{equation}
\label{epsiloneffeomtphi}
\left[\tilde{\square}_{{\rm AdS}_2}-\bar{\mu}^2\right]\phi_\ell=\mathcal{O}\left(\epsilon\right),
\end{equation}
where $\tilde{\square}_{{\rm AdS}_2}=2\partial_{\tilde{v}}\partial_{\tilde{\rho}}+\partial_{\tilde{\rho}}(\lambda_0 \tilde{\rho}^2\partial_{\tilde{\rho}})$. We have used a property ${\square}_{{\rm AdS}_2}\to\tilde{\square}_{{\rm AdS}_2}$ because of the scaling symmetry of ${\rm AdS}_2$. In the limit of $\epsilon\to0$, we can obtain late-time asymptotic behaviors of $\phi_\ell$ near the horizon. 

In the limit of $\epsilon\to0$, Eq.~\eqref{epsiloneffeomtphi} becomes
\begin{equation}
\label{eomtphi0}
\left[\tilde{\square}_{{\rm AdS}_2}-\bar{\mu}^2\right]\phi_\ell=0.
\end{equation}
This is exactly the massive Klein-Gordon equation with mass squared~$\bar{\mu}^2$ in ${\rm AdS}_2$. Hence, solutions of Eq.~\eqref{epsiloneffeomtphi} can be written in the form
\begin{equation}
\label{epsilonexpansionphi}
\phi_\ell\left(\tilde{v},\tilde{\rho}\right)=\phi_\ell^{{\rm AdS}_2}\left(\tilde{v},\tilde{\rho}\right)+\mathcal{O}(\epsilon),
\end{equation}
where $\phi_\ell^{{\rm AdS}_2}\left(\tilde{v},\tilde{\rho}\right)$ satisfies Eq.~\eqref{eomtphi0}. Thus, $\phi_\ell$ is in good agreement with $\phi_\ell^{{\rm AdS}_2}$ at the late time $\epsilon\to0$.

\section{Late-time behaviors of the massive Klein-Gordon field near the extremal black hole horizon and horizon-instability}
\label{sec:latetimepowerlawtailsandinstability}
We discuss late-time behaviors of the massive Klein-Gordon field that satisfies Eq.~\eqref{KGE2} near the horizon in terms of the near-horizon geometry. As shown in the previous section, the study of this topic is reduced to the analysis of the massive scalar fields in ${\rm AdS}_2$, i.e., Eq.~\eqref{eomtphi0}. Although there is a heuristic argument in~\cite{Lucietti:2012xr}, we revisit this problem by analyzing the general normalizable solutions in ${\rm AdS}_2$. We give the detailed calculations in Appendix~\ref{appendix:MKGfiedsinAdS2} and summarize the results in this section.

\subsection{Specific mass squared case: $\mu^2=\lambda_0N(N+1)-\ell(\ell+n-3)/\gamma_0$}
We discuss the cases where the Aretakis constants exist. Analysis in Appendix~\ref{appendix:MKGfiedsinAdS2} implies that the late-time behavior of the field near the horizon takes a power law of time,
\begin{equation}
\label{latetimetail}
\phi_\ell\propto \left(-\frac{v}{2}-\frac{\lambda_0\rho v^2}{4}\right)^{-N-1}.
\end{equation}
We notice that Eq.~\eqref{latetimetail} implies that $\partial_\rho^{k(\le N)}\phi_\ell|_{\rho=0}$ all decay at late times and $\partial_\rho^{N+1}\phi_\ell|_{\rho=0}$ is constant. In other words,  the Aretakis constants~\eqref{garec} take the form
\begin{equation}
\label{latetimearec}
\mathcal{H}_N\simeq \partial_\rho^{N+1}\phi_\ell\big|_{\rho=0},~~{\rm as}~v\to\infty.
\end{equation}
Equation~\eqref{latetimearec} provides the pre-factor of the leading asymptotic behavior~\eqref{latetimetail}:  
\begin{equation}
\label{latetimeouterthehorizon}
\phi_\ell\simeq \frac{N!}{\lambda_0^{N+1}\left(2N+1\right)!}\mathcal{H}_{N}\left(-\frac{v}{2}-\frac{\lambda_0\rho v^2}{4}\right)^{-N-1}.
\end{equation}
This is consistent with the results of~\cite{Aretakis:2011hc,Lucietti:2012xr,Aretakis:2011ha,Murata:2012ct,Bhattacharjee:2018pqb,Ori:2013iua,Sela:2015vua,Angelopoulos:2018uwb,Blaksley:2007ak}. The late-time tails~\eqref{latetimeouterthehorizon} also imply that $\partial_\rho^{k(\ge N+2)}\phi_\ell|_{\rho=0}$ blow up polynomially in $v$, i.e.,
\begin{equation}
\label{divergentbehavior1}
\partial_{\rho}^{k}\phi_\ell\big|_{\rho=0}\simeq \frac{(N+k)!}{2^{k-N-1}(2N+1)!}\mathcal{H}_N \left(-\frac{1}{\lambda_0}\right)^{N+1-k} v^{k-N-1},~~{\rm as}~~v\to \infty~(k\ge N+2).
\end{equation}
This is the Aretakis instability. In the present case, the divergent behaviors can only occur in the second-order or higher derivatives. 

\subsection{General mass squared cases with $\bar{\mu}^2\ge -\lambda_0/4$}
In this case, the Aretakis constants do not necessarily exist in general. We assume that the effective mass squared~$\bar{\mu}^2$ in Eq.~\eqref{barmu} satisfies $\bar{\mu}^2/\lambda_0\ge m^2_{{\rm BF},2}$, where $m^2_{{\rm BF},2}=-1/4$ is the Breitenlohner-Freedman (BF) bound in ${\rm AdS}_2$~\cite{Breitenlohner:1982jf,Breitenlohner:1982bm}. It is known that without this assumption, exponentially growing unstable modes appear~\cite{Breitenlohner:1982jf,Breitenlohner:1982bm}. Analysis in terms of the near-horizon geometry in Appendix~\ref{appendix:MKGfiedsinAdS2} implies that the leading late-time behavior is described by power-law tails
\begin{equation}
\label{latetimeouterthehorizon2}
\phi_\ell\simeq \frac{\Gamma(\Delta)}{\lambda_0^\Delta\Gamma(2\Delta)}H_\Delta\left(-\frac{v}{2}-\frac{\lambda_0\rho v^2}{4}\right)^{-\Delta},
\end{equation}
where $H_\Delta$ is a constant. This shows that for $k\ge \lfloor \Delta\rfloor+1$, where $\lfloor \, \rfloor$ denotes the integer part, $k$th-derivatives of $\phi_\ell$ at the horizon $\rho=0$ blow up polynomially in $v$ at the late time, i.e.,
\begin{equation}
\label{divergentbehavior2}
\partial_{\rho}^{k}\phi_\ell\big|_{\rho=0}\simeq \frac{\Gamma(\Delta+k)}{2^{k-\Delta}\Gamma(2\Delta)}H_\Delta \left(-\frac{1}{\lambda_0}\right)^{\Delta-k} v^{k-\Delta}.
\end{equation}
This is consistent with the results in \cite{Lucietti:2012xr}. 

We notice that $\partial_\rho\phi_\ell|_{\rho=0}$ is divergent at the late time if $\lfloor \Delta\rfloor=0$ or equivalently the mass squared $\mu^2$ is in the range
\begin{equation}
\label{massrange}
-\frac{1}{4}\le\frac{\mu^2}{\lambda_0}+\frac{\ell(\ell+n-3)}{\lambda_0\gamma_0}<0.
\end{equation}
The blowup of the first-order derivative directly implies that of the physical quantity at the horizon. Namely, defining the energy-momentum tensor of the massive Klein-Gordon field~$\Phi$ satisfying Eq.~\eqref{KGE2},
\begin{equation}
T_{\mu\nu}=\nabla_\mu\Phi\nabla_\nu\Phi-\frac{1}{2}g_{\mu\nu}\left[\nabla_\sigma\Phi\nabla^\sigma\Phi+\mu^2\Phi\right],
\end{equation}
we can see that the energy density measured by an infalling observer at the horizon, $T_{\rho\rho}$, is divergent at the late time for the mass squared $\mu^2$ in the range~\eqref{massrange}.

We note that the asymptotic structure of the spacetime in the far region is not specified. If one considers an $n$-dimensional asymptotically AdS extremal black hole, slightly negative mass squared of a Klein-Gordon field is allowed for the stable dynamical evolution in the asymptotic region, i.e., $\mu^2/\lambda_0\ge m^2_{\rm BF,n}=-(n(n-2)+1)/4$, where $ m^2_{\rm BF,n}$ is the BF bound in ${\rm AdS}_n$~\cite{Breitenlohner:1982jf,Breitenlohner:1982bm}. 
The BF bound in ${\rm AdS}_2$ is effectively violated in the near-horizon geometry if the massive Klein-Gordon field with $\ell=0$ in $n$-dimensions has mass squared of $m^2_{{\rm BF},n}\le \mu^2/\lambda_0<-1/4=m^2_{{\rm BF},2}$. This violation is discussed in the context of the holographic superconductor~\cite{Hartnoll:2008kx}. While one may think that physical quantities do not blow up without the violation of the BF bound, our result shows that extremal black holes suffer from blowup of components of the energy-momentum tensor of the test field at the horizon if the mass squared  satisfies the inequality~\eqref{massrange}.

\section{Aretakis constants and instability in the parallelly propagated null geodesic frame} 
\label{sec:AretakisconstandinstabilityinPP}
We discuss geometrical meanings of the Aretakis constants~\eqref{garec} and 
instability~\eqref{divergentbehavior2} in terms of the parallelly propagated null geodesic frame along the extremal black hole horizons. We shall explicitly show that the Aretakis constants and instability correspond to, respectively, constants and blowups of some components of higher-order covariant derivatives of the field at late times in that frame along the horizons. This shows that a similar geometrical interpretation as in the ${\rm AdS}_2$ case~\cite{Katagiri:2021scx} is possible for our present setup.

We introduce vector fields in the extremal black hole spacetime~\eqref{EBH},
\begin{equation}
\begin{split}
&e_{(0)}^\mu \partial_\mu = \partial_v,~e_{(1)}^\mu \partial_\mu = -\partial_\rho,~e_{(A)}^\mu \partial_\mu = \sqrt{g^{\theta^A\theta^A}}\partial_{\theta^A}.
\end{split}
\end{equation}
At the horizon $\rho=0$, these satisfy 
\begin{equation}
\begin{split}
\label{basisrelation1}
& e_{(0)}^\mu \nabla_\mu e_{(0)}^\nu = 0,~ 
e_{(0)}^\mu \nabla_\mu e_{(1)}^\nu = 0,~e_{(0)}^\mu \nabla_\mu e_{(A)}^\nu = 0.
\\
& e_{(0)}^\mu e_{(0)\mu}= 0,~ 
e_{(1)}^\mu e_{(1)\mu} = 0,~e_{(A)}^\mu e_{(B)\mu} = \delta_{AB},\\
&e_{(0)}^\mu e_{(1)\mu} = -1,~e_{(0)}^\mu e_{(A)\mu} = 0,~e_{(1)}^\mu e_{(A)\mu} = 0.
\end{split}
\end{equation}
Here, $A,B=2,3,\cdots, n-1$. The vectors~$e_{(0)}^\mu,e_{(1)}^\mu,e_{(A)}^\mu$ are parallelly transported along a null geodesic $e_{(0)}^\mu$ at the horizon $\rho=0$. The frame formed by them is called the {\it parallelly propagated null geodesic frame} along the horizon.

We find the relation
\begin{equation}
\begin{split}
\left.\left(-1\right)^{i}e_{(1)}^{\mu_1}e_{(1)}^{\mu_2} \cdots e_{(1)}^{\mu_{i}}
\nabla_{\mu_1}\nabla_{\mu_2} \cdots \nabla_{\mu_{i}} \phi_\ell\right|_{\rho = 0} = 
\left.
\partial_{\rho}^{i} \phi_\ell
\right|_{\rho = 0},
\label{e1e1nablanabla}
\end{split}
\end{equation}
where $i$ is a nonnegative integer. For the specific cases~$\mu^2=\lambda_0 N(N+1)-\ell(\ell+n-3)/\gamma_0$, in which the Aretakis constants exist, combining Eqs.~\eqref{latetimearec} and~\eqref{e1e1nablanabla}, the relation
\begin{equation}
\begin{split}
\left.\left(-1\right)^{N+1}e_{(1)}^{\nu_1}e_{(1)}^{\nu_2} \cdots e_{(1)}^{\nu_{N+1}}
\nabla_{\nu_1}\nabla_{\nu_2} \cdots \nabla_{\nu_{N+1}} \phi_\ell\right|_{\rho = 0}\simeq \mathcal{H}_{N},
\end{split}
\end{equation}
holds at the late time~$v\to\infty$. The Aretakis constants correspond to the late-time behavior of the component of $(N+1)$th-order covariant derivatives of the field at the late time in the parallelly propagated null geodesic frame along the horizon. 

Equation~\eqref{e1e1nablanabla} also shows that the divergent behavior~\eqref{divergentbehavior2} for $k=\lfloor \Delta\rfloor+1$ corresponds to that of  $(\lfloor \Delta\rfloor+1)$th-order covariant derivatives at the late time in that frame along the horizon:
\begin{equation}
\begin{split}
\left.\left(-1\right)^{\lfloor \Delta\rfloor+1}e_{(1)}^{\nu_1}e_{(1)}^{\nu_2} \cdots e_{(1)}^{\nu_{\lfloor \Delta\rfloor+1}}
\nabla_{\nu_1}\nabla_{\nu_2} \cdots \nabla_{\nu_{\lfloor \Delta\rfloor+1}} \phi_\ell\right|_{\rho = 0}\sim v.
\label{e1e1nablanabla2}
\end{split}
\end{equation} 
In the same manner, it can be shown that the $(\lfloor \Delta\rfloor+2)$th-and higher-order covariant derivatives are also unbounded.\footnote{Other components of the $\left(\lfloor \Delta\rfloor+1\right)$th-order covariant derivatives at the horizon are bounded: we have a relation,
\begin{equation}
\begin{split}
\left(-1\right)^{\lfloor \Delta\rfloor+1-j}e_{(0)}^{\nu_1} \cdots e_{(0)}^{\nu_{p}}
e_{(1)}^{\nu_{j+1}} \cdots e_{(1)}^{\nu_{\lfloor \Delta\rfloor+1-j}}\left.\nabla_{\nu_1}\cdots \nabla_{\nu_{j}} 
\nabla_{\nu_{j+1}} \cdots \nabla_{\nu_{\lfloor \Delta\rfloor+1}} 
\phi_\ell\right|_{\rho = 0}= 
\left.\partial_v^{j} \partial_{\rho}^{\lfloor \Delta\rfloor+1-j} \phi_\ell\right|_{\rho = 0},
\label{e0e1ppframe}
\end{split}
\end{equation}
for a positive integer $j$. We note that components including $e_{(A)}^{\mu}$ vanish. The right-hand side vanishes or decays at the late time because $\partial_\rho^{\lfloor \Delta\rfloor+1-j}\phi_\ell|_{\rho = 0}$ are constant or decay. Hence, the above relation shows that all of the components with respect to $e_{(0)}^{\mu_1} \cdots e_{(0)}^{\mu_{p}}e_{(1)}^{\mu_{p+1}} \cdots e_{(1)}^{\mu_{\lfloor \Delta\rfloor+1-j}}$ are bounded. Likewise, all of the components of the lower order covariant derivatives are bounded.} The Aretakis instability corresponds to blowups of components of the covariant derivatives of the fields at the late time in the parallelly propagated null geodesic frame along the horizons.

\section{Aretakis constants from ladder operators associated with approximate spacetime conformal symmetries}
\label{sec:Aretakisconstsfromladder}
According to~\cite{Cardoso:2017qmj,Katagiri:2021scx}, the Aretakis constants in ${\rm AdS}_2$ can be derived by using ladder operators, called mass ladder operators~\cite{Cardoso:2017qmj,Katagiri:2021scx,Cardoso:2017egd}, constructed from the spacetime conformal symmetry. Since ${\rm AdS}_2$ structure approximately appear in the vicinity of extremal black hole horizons, we expect that the Aretakis constants~\eqref{garec} can also be derived similarly as~\cite{Cardoso:2017qmj,Katagiri:2021scx}.  From this point of view, we construct the Aretakis constants in the extremal black holes~\eqref{EBH} in this section.

\subsection{Mass ladder operators in ${\rm AdS}_2$}
\label{subsec:massladderinAdS2}
We first review the {\it mass ladder operator} in ${\rm AdS}_2$~\cite{Cardoso:2017qmj,Katagiri:2021scx,Cardoso:2017egd}. The mass ladder operator can be defined in spacetimes with a spacetime conformal symmetry, e.g., the AdS spacetime. In the Eddington-Finkelstein coordinates~$\left(v,\rho\right)$, which cover the future Poincar\'e horizon of ${\rm AdS}_2$ located at $\rho=0$, the line element is given by
\begin{equation}
\label{asym}
ds^2=-\lambda_0 \rho^2dv^2+2dvd\rho,
\end{equation}
where $\lambda_0$ is a positive constant. The constant~$\lambda_0$ is associated with the absolute value of the scalar curvature of ${\rm AdS}_2$: $R_{{\rm AdS}_2}=-2\lambda_0$. 

We define the mass ladder operators in ${\rm AdS}_2$, which act on scalar fields, as
\begin{equation}
\label{mlo}
D_{k}:=\mathcal{L}_{\zeta}-\frac{k}{2}\nabla_\mu\zeta^\mu,~~k\in\mathbb{R},
\end{equation}
where $\zeta^\mu$ is a closed conformal Killing vector in ${\rm AdS}_2$, which satisfies the closed conformal Killing equation
\begin{equation}
\nabla_\mu\zeta_\nu=\frac{1}{2}\nabla_\sigma \zeta^\sigma g_{\mu\nu}.
\end{equation}
For this operator, we have a commutation relation
\begin{equation}
[\square_{{\rm AdS}_2}, D_{k}] = -2\lambda_0  k D_{k} +(\nabla_\mu \zeta^\mu)\left[\Box_{{\rm AdS}_2} -\lambda_0 k (k+1)\right],
\label{eq:comrelformassladder}
\end{equation}
where $\square_{{\rm AdS}_2}:= 2\partial_v\partial_\rho+\partial_\rho(\lambda_0\rho^2\partial_\rho)$ is defined in Eq.~\eqref{BoxAdS2}. This commutation relation can be written in the form
\begin{equation}
\label{relatiomassladder}
D_{k-2}\left[\square_{{\rm AdS}_2}-k(k+1)\right]=\left[\square_{{\rm AdS}_2}-k(k-1)\right]D_{k}.
\end{equation}
Acting Eq.~\eqref{relatiomassladder} on the massive Klein-Gordon field $\phi$ with mass squared $m^2 =k(k+1)$, we find that $D_k\phi$ satisfies the massive Klein-Gordon equation with mass squared $m^2 = k(k-1)$. The mass ladder operators map a solution of the massive Klein-Gordon equation to that of the Klein-Gordon equation with different mass squared.

${\rm AdS}_2$ admits three independent closed conformal Killing vectors, but in this paper we focus on the specific one
\begin{equation}
\begin{split}
\label{ckvads}
\zeta&=v^2\partial_v+\left(\frac{2}{\lambda_0 }+2v \rho+\lambda_0 v^2\rho^2\right)\partial_\rho.
\end{split}
\end{equation}
For this closed conformal Killing vector, the mass ladder operator in Eq.~\eqref{mlo} becomes
\begin{equation}
\begin{split}
\label{mloinAdS2}
D_{k}&=v^2\partial_v+\left(\frac{2}{\lambda_0 }+2v \rho+\lambda_0 v^2\rho^2\right)\partial_\rho-kv(2+\lambda_0 v\rho).
\end{split}
\end{equation}

\subsection{Aretakis constants in ${\rm AdS}_2$ from the spacetime conformal symmetry}
\label{subsec:AretakisconstsinAdS2}
We explain that the Aretakis constants in ${\rm AdS}_2$ can be constructed from the mass ladder operators~\eqref{mloinAdS2} based on~\cite{Cardoso:2017qmj,Katagiri:2021scx}. Let us consider massive Klein-Gordon fields $\phi(v,\rho)$ with mass squared $m^2=\lambda_0 N(N+1)$ $(N=0,1,2,\cdots)$ satisfying
\begin{equation}
\label{KGEQAdS2}
\left[\square_{{\rm AdS}_2} - \lambda_0 N(N+1)\right]\phi= 0.
\end{equation}
We note here again that $\lambda_0 $ is positive and real. In ${\rm AdS}_2$, quantities defined by
\begin{equation}
\label{arecinAdS2}
\mathcal{H}_N^{{\rm AdS}_2}=\partial_\rho^{N+1}\phi|_{\rho=0},
\end{equation}
are independent of $v$ on the horizon $\rho=0$. The quantities~$\mathcal{H}_N^{{\rm AdS}_2}$ are called the Aretakis constants in ${\rm AdS}_2$~\cite{Lucietti:2012xr,Cardoso:2017qmj,Katagiri:2021scx}. First, we consider the massless case $N= 0$. Then, the Klein-Gordon equation \eqref{KGEQAdS2} can be written as
\begin{equation}
\label{Rmasslesseom}
2 \partial_v \partial_\rho \phi = -\lambda_0 \rho \left(2\partial_\rho + \rho \partial_\rho^2\right)\phi.
\end{equation}
Evaluating this at the  horizon $\rho = 0$, we can see $\partial_v\mathcal{H}_0^{{\rm AdS}_2}=0$. Thus, the Aretakis constant in ${\rm AdS}_2$ is derived for $N=0$ case.

Next, we consider $N\ge1$ cases. Using the commutation relations~\eqref{eq:comrelformassladder}, we can show
\begin{equation}
\begin{split}
\label{massless}
D_{-1}D_{0}\cdots D_{N-2}\left[\Box_{{\rm AdS}_2}-\lambda_0 N(N+1)\right]\phi=\Box_{{\rm AdS}_2}D_{1}D_{2}\cdots D_{N}\phi
\end{split}
\end{equation}
Since the left-hand side vanishes due to the massive Klein-Gordon equation for $\phi$, this yields
\begin{equation}
\begin{split}
\label{massless_eom}
\Box_{{\rm AdS}_2}D_{1}D_{2}\cdots D_{N}\phi=0.
\end{split}
\end{equation}
It follows that $D_{1}D_{2}\cdots D_N\phi$ satisfies the massless Klein-Gordon equation. Thus, solutions of the massive Klein-Gordon equation with the mass squared $m^2 =\lambda_0  N(N +1)$ in ${\rm AdS}_2$ can be mapped into that of the massless Klein-Gordon equation. Note that this is not the case for other parameterization of mass squared. As in the massless case, Eq.~\eqref{massless_eom} can be rewritten as
\begin{equation}
2 \partial_v \partial_\rho D_{1}D_{2}\cdots D_{N}\phi= -\lambda_0 \rho \left(2\partial_\rho + \rho \partial_\rho^2\right)D_{1}D_{2}\cdots D_{N}\phi.
\end{equation}
This is the same form as Eq.~\eqref{Rmasslesseom}. Thus, there exist conserved quantities along the future Poincar\'e horizon, which are defined by
\begin{align}
\label{QAdS2}
\left.{\cal Q}_N ^{{\rm AdS}_2}:= \partial_\rho D_{1}D_{2}\cdots D_{N}\phi\right|_{\rho=0}.
\end{align}
It has been shown that $\left({\lambda_0 }/{2}\right)^{N}{\cal Q}_N^{{\rm AdS}_2}$ coincide with the Aretakis constants $\mathcal{H}_N^{{\rm AdS}_2}$ in Eq.~\eqref{arecinAdS2}~\cite{Cardoso:2017qmj,Katagiri:2021scx}.

\subsection{The Aretakis constants in the extremal black holes from the approximate spacetime conformal symmetry}
\label{subsec:Aretakisconstsfromladder-B}
We apply the previous manner to the case for the Aretakis constants~\eqref{arec} in the extremal black hole spacetimes~\eqref{EBH}. 
As seen in Sec.~\ref{subsec:MKGfiedsinNHG}, the Klein-Gordon equation~\eqref{eom2} in the vicinity of the extremal black hole horizons can effectively be described by the massive Klein-Gordon field in ${\rm AdS}_2$ as
\begin{equation}
\label{effeommain1}
\left[\square_{{\rm AdS}_2}-\mu^2-\frac{\ell(\ell+n-3)}{\gamma_0}\right]\phi_\ell=\delta\left[\phi_\ell\right],
\end{equation}
where $\square_{{\rm AdS}_2}$ and $\delta[\phi_\ell]$ are defined in Eq.~\eqref{BoxAdS2} and Eq.~\eqref{deltaN}, respectively. For the case~$\mu^2=\lambda_0 N(N+1)-\ell(\ell+n-3)/\gamma_0$, where the Aretakis constants exist, Eq.~\eqref{effeommain1} becomes
\begin{equation}
\label{effeommain}
\left[\square_{{\rm AdS}_2}-\lambda_0N(N+1)\right]\phi_\ell=\delta\left[\phi_\ell\right].
\end{equation}
Hereafter, we consider this equation.

For later convenience, we note that $\delta[\phi_\ell]$ can be written in the form
\begin{equation}
\label{relationP}
\partial_{\rho}^{N}\delta[\phi_\ell]=\partial_vP_N+\mathcal{O}\left(\rho\right),
\end{equation}
where
\begin{equation}
\begin{split}
\label{functionP}
\gamma_0P_N(v)=&-\left.\partial_\rho^{N}\left[2\delta\gamma\partial_\rho\phi_\ell+\left(\partial_\rho\delta\gamma\right)\phi_\ell\right]\right|_{\rho=0}\\
&-\sum_{i=0}^{N+1}\frac{(N+1)!}{i!(N+1-i)!}\left.\partial_\rho^i\left[\rho^2\left(\gamma_0\delta\lambda+\lambda_0 \delta\gamma+\delta\lambda\delta\gamma\right)\right]\right|_{\rho=0}\mathcal{G}_{N+2-i}(v)\\
&+\mu^2\sum_{i=0}^{N}\frac{N!}{(N-i)!}\gamma_i\mathcal{G}_{N-i}(v),
\end{split}
\end{equation}
where $\gamma_i$ is given in Eq.~\eqref{labellambdagamma}. Here, we have defined
\begin{equation}
\label{Gj}
\left.\mathcal{G}_{N-j}(v):=\frac{2}{\lambda_0 \gamma_0j(2N-j+1)}\partial_{\rho}^{N-j}\left[\left(\gamma_0+\delta\gamma\right)^{1/2}\partial_\rho\left\{\left(\gamma_0+\delta\gamma\right)^{1/2}\phi_\ell\right\}\right]\right|_{\rho=0}.
\end{equation}
The derivation of Eq.~\eqref{relationP} is in Appendix~\ref{Appendix:P}. Note that for the cace $\gamma_0\delta\lambda+\lambda_0 \delta\gamma+\delta\lambda\delta\gamma=0$, e.g., the four-dimensional extremal Reissner-Nordstr\"om spacetime, the second line of the right-hand side in Eq.~\eqref{functionP} vanishes.

Using the commutation relations~\eqref{eq:comrelformassladder} on Eq.~\eqref{effeommain}, we can rewrite the left-hand side to
\begin{equation}
\begin{split}
\label{massless_eom2}
D_{-1}D_{0}\cdots D_{N-2}\left[\Box_{{\rm AdS}_2}-\lambda_0 N(N+1)\right]\phi_\ell=\Box_{{\rm AdS}_2}D_{1}D_{2}\cdots D_{N}\phi_\ell,
\end{split}
\end{equation}
where $D_k$ is the differential operator whose form is Eq.~\eqref{mloinAdS2}. This is equivalent to the relation~\eqref{massless}. Using Eq.~\eqref{massless_eom2}, Eq.~\eqref{effeommain} is rewritten as
\begin{equation}
\begin{split}
\label{redl}
\Box_{{\rm AdS}_2}D_{1}D_{2}\cdots D_{N}\phi_\ell=D_{-1}D_{0}\cdots D_{N-2}\delta[\phi_\ell].
\end{split}
\end{equation}
As shown in Appendix~\ref{Appendix:B}, the right-hand side of Eq.~\eqref{redl} has the relation
\begin{equation}
\label{DDDdelta}
D_{-1}D_{0}\cdots D_{N-2}\delta\left[\phi_\ell\right]=-\partial_v\Lambda_N+\mathcal{O}(\rho),
\end{equation}
where
\begin{equation}
\begin{split}
\label{LambdaN}
-\Lambda_N=&v^2D_{0}D_{1}\cdots D_{N-1}D_{N-2}\delta\left[\phi_\ell\right]+\left(\frac{2}{\lambda_0 }\right)v^2\partial_\rho D_{1}D_{2}\cdots D_{N-1}D_{N-2}\delta\left[\phi_\ell\right]+\\
&\cdots+\left(\frac{2}{\lambda_0 }\right)^{N-2}v^2\partial_\rho^{N-2}D_{N-2}\delta\left[\phi_\ell\right]+\left(\frac{2}{\lambda_0 }\right)^{N-1}v^2\partial_\rho^{N-1}\delta\left[\phi_\ell\right]+\left(\frac{2}{\lambda_0 }\right)^{N}P_N.
\end{split}
\end{equation}
With this relation, Eq.~\eqref{redl} is written as
\begin{equation}
\begin{split}
\Box_{{\rm AdS}_2}D_{1}D_{2}\cdots D_{N}\phi_\ell=-\partial_v\Lambda_N+\mathcal{O}(\rho).
\end{split}
\end{equation}
Furthermore, using the explicit form of $\Box_{{\rm AdS}_2}$ in Eq.~\eqref{BoxAdS2}, this equation can be rewritten as
\begin{equation}
\begin{split}
\label{conservationlaw}
&2\partial_v\left[ \partial_\rho D_{1}D_{2}\cdots D_{N}\phi_\ell+\frac{1}{2}\Lambda_N\right]=-\lambda_0 \partial_\rho\left[\rho^2\partial_\rho\left(D_{1}D_{2}\cdots D_{N}\phi_\ell\right)\right]+\mathcal{O}(\rho).
\end{split}
\end{equation}
Because the first term of the right-hand side of Eq.~\eqref{conservationlaw} is also $\mathcal{O}(\rho)$, at the horizon $\rho=0$ Eq.~\eqref{conservationlaw} shows
\begin{equation}
\begin{split}
\label{newconl}
\partial_v\mathcal{Q}_N=0,
\end{split}
\end{equation}
where
\begin{equation}
\begin{split}
\label{QN}
\mathcal{Q}_N:=\partial_\rho D_{1}D_{2}\cdots D_{N}\phi_\ell\big|_{\rho=0}+\frac{1}{2}\Lambda_N.
\end{split}
\end{equation}
It follows from Eq.~\eqref{newconl} that $\mathcal{Q}_N$ are independent of $v$ at the black hole horizon~$\rho=0$. Hence, $\mathcal{Q}_N$ are conserved along the horizon in the arbitrary dimensional extremal black holes~\eqref{EBH}. 
As shown below, we show that the leading late-time contribution of $\mathcal{Q}_N$ is the first term of Eq.~\eqref{QN},
and $(\lambda_0 /2)^{N}\mathcal{Q}_N$ asymptotes to the Aretakis constant~\eqref{garec} (see also Eq.~\eqref{latetimearec}). This implies that $(\lambda_0 /2)^{N}\mathcal{Q}_N$ coincides with the Aretakis constant~\eqref{garec} everywhere in the horizon $\rho = 0$.\footnote{The explicit calculation can also show this result. The explicit examples of the cases for $N = 1, 2$ are given in  Appendix~\ref{appendix:Examples}.} The quantities $(\lambda_0 /2)^{N}\partial_\rho D_{1}D_{2}\cdots D_{N}\phi_\ell|_{\rho=0}$, which correspond to the Aretakis constants in ${\rm AdS}_2$ constructed from the ${\rm AdS}_2$ conformal symmetry, asymptote to the Aretakis constants~\eqref{garec} at the late time.
This is the relation between the Aretakis constants of the extremal black boles and the approximate ladder operators constructed from the approximate spacetime conformal symmetry.\footnote{As shown in~\cite{Katagiri:2021scx}, in the case of ${\rm AdS}_2$, the same quantity as Eq.~\eqref{QAdS2} constructed from the closed conformal Killing vector, which is null at the horizon is proportional to the multiplication of the positive power of $\rho$ and the Aretakis constant, and then it vanishes at the horizon. Thus, in the case of extremal black holes, we expect that the similar conserved quantity vanishes if the corresponding approximate conformal Killing vector is null at the horizon. This is the reason why the discussion in~\cite{Cardoso:2017qmj} does not work as mentioned in footnote~\ref{footnote:intro2}. In this paper, we focus on the approximate conformal Killing vector whose form is given in Eq.~\eqref{ckvads} and this is spacelike at the horizon, and the corresponding conserved quantities do not vanish.}

\subsection{Late-time expressions for the quantities $\mathcal{Q}_N$}
\label{subsec:QNinepsilon}
Finally, we discuss late-time expressions for the quantities $\mathcal{Q}_N$ in Eq.~\eqref{QN}. We expect that $\mathcal{Q}_N$ approach $\mathcal{Q}_N^{{\rm AdS}_2}$ in Eq.~\eqref{QAdS2} in $v\to\infty$ because the AdS structure is a good approximation to the neighborhood of extremal black holes at late times as seen in Sec.~\ref{subsec:MKGfiedsinNHG}. This is also expected from the fact that the Aretakis constants in the present system are expressed by those in ${\rm AdS_2}$ at the late time as seen in Eq.~\eqref{latetimearec}. To check that, we investigate properties of $\mathcal{Q}_N$ for the coordinate transformation $(v,\rho)\to(\tilde{v}/\epsilon,\epsilon\tilde{\rho})$. For the transformation, we have the relations
\begin{equation}
\label{operatorrelationsforepsilon}
\partial_v\to \epsilon \partial_{\tilde{v}},~~\partial_\rho\to \epsilon^{-1} \partial_{\tilde{\rho}},~~D_{k}\to \epsilon^{-1}\tilde{D}_k,
\end{equation}
where $\tilde{D}_k=\tilde{v}^2\partial_{\tilde{v}}+({2}/{\lambda_0 }+2\tilde{v} \tilde{\rho}+\lambda_0 \tilde{v}^2\tilde{\rho}^2)\partial_{\tilde{\rho}}-k\tilde{v}(2+\lambda_0 \tilde{v}\tilde{\rho})$.

We first focus on $\partial_\rho D_{1}D_{2}\cdots D_{N}\phi_\ell|_{\rho=0}$ in $\mathcal{Q}_N$. As seen in Sec.~\ref{subsec:MKGfiedsinNHG}, $\phi_\ell(\tilde{v}/\epsilon,\epsilon \tilde{\rho})$ can be expanded in $\epsilon$ as
\begin{equation}
\label{epsilonexpansionphiell}
\phi_\ell\left(\tilde{v}/\epsilon,\epsilon\tilde{\rho}\right)=\phi_\ell^{{\rm AdS}_2}\left(\tilde{v}/\epsilon,\epsilon\tilde{\rho}\right)+\mathcal{O}(\epsilon),
\end{equation}
where $\phi_\ell^{{\rm AdS}_2}$ satisfies the massive Klein-Gordon equation in ${\rm AdS}_2$. Using the relation~\eqref{operatorrelationsforepsilon}, Eq.~\eqref{epsilonexpansionphiell} implies
\begin{equation}
\partial_\rho D_{1}D_{2}\cdots D_{N}\phi_\ell\big|_{\rho=0}\to \epsilon^{-N-1}\left(\mathcal{Q}_N^{{\rm AdS}_2}+\mathcal{O}\left(\epsilon\right)\right).
\end{equation}
Next, noticing $\delta[\phi_\ell]=\mathcal{O}(\epsilon)$, we see from Eq.~\eqref{relationP} that $P_N=\mathcal{O}(\epsilon^{-N})$, and hence
\begin{equation}
\Lambda_N=\mathcal{O}(\epsilon^{-N}),
\end{equation}
from Eq.~\eqref{LambdaN}. Equation~\eqref{newconl} thus shows 
\begin{equation}
\partial_{\tilde{v}}\left[\mathcal{Q}_N^{{\rm AdS}_2}+\mathcal{O}\left(\epsilon\right)\right]=0.
\end{equation}
Hence, $\mathcal{Q}_N$ behave as $\mathcal{Q}_N^{{\rm AdS}_2}+\mathcal{O}(\epsilon)$ for the transformation $(v,\rho)\to(\tilde{v}/\epsilon,\epsilon\tilde{\rho})$. We can indeed see that $\mathcal{Q}_N$ approach $\mathcal{Q}_N^{{\rm AdS}_2}$ at the late time $\epsilon\to 0$. 

\section{Summary and discussion}
In this paper, we have investigated late-time behaviors of massive Klein-Gordon fields in general static and spherically symmetric extremal black hole spacetimes in arbitrary dimensions. We have shown that there exist conserved quantities along the extremal black hole horizons, which are the Aretakis constants, if the quantity
\begin{equation}
\label{DeltaVI}
\Delta=\frac{1}{2}+\sqrt{\frac{\mu^2}{\lambda_0 }+\frac{\ell(\ell+n-3)}{\lambda_0 \gamma_0}+\frac{1}{4}},
\end{equation}
is an integer larger than or equal to unity, where $\mu^2$, $\ell$, and $n$ are the mass squared and multipole number of the scalar fields, and the number of spacetime dimensions, respectively. In Eq.~\eqref{DeltaVI}, the constants~$\lambda_0$ and $\gamma_0$ are quantities associated with the near-horizon geometry which is described by $ {\rm AdS}_2 \times S^{n-2}$ with the effective ${\rm AdS_2}$ radius $\lambda_0^{-1/2}$ and the horizon area radius $\gamma_0^{1/2}$.
For example, the Aretakis constants exist for a massless scalar field in the arbitrary dimensional extremal Reissner-Nordstr\"om spacetime, and a scalar field with $\ell=0$ and specific mass squared~$\mu^2 = \lambda_0N(N + 1)$ for nonnegative integers~$N$ in general static and spherically symmetric extremal black holes in arbitrary dimensions. We have also derived late-time behaviors of the fields near the horizon, i.e., power-law tails, in terms of the near-horizon geometry. The late-time power-law tails lead to the Aretakis instability: the polynomial growth of the higher-order transverse derivatives of the fields on the horizon at the late time. We have checked that our results are consistent with the previous works~\cite{Aretakis:2011hc,Lucietti:2012xr,Aretakis:2011ha,Murata:2012ct,Bhattacharjee:2018pqb,Ori:2013iua,Sela:2015vua,Angelopoulos:2018uwb,Lucietti:2012sf,Katagiri:2021scx,Blaksley:2007ak}.

We have also discussed geometrical meanings of the Aretakis constants and instability. 
We have demonstrated that the Aretakis constants and instability correspond to respectively constants and blowups of components of covariant derivatives of the field at late times in the parallelly propagated null geodesic frame along the horizons. Furthermore, we have derived the Aretakis constants by the mass ladder operators constructed from approximate ${\rm AdS}_2$ symmetry~\cite{Cardoso:2017qmj,Katagiri:2021scx,Cardoso:2017egd}. 

If the effective mass squared~$\bar{\mu}^2$ defined in Eq.~\eqref{barmu} satisfies $\bar{\mu}^2/\lambda_0<m^2_{{\rm BF},2}$, where $m^2_{{\rm BF},2}=-1/4$ is the Breitenlohner-Freedman (BF) bound in ${\rm AdS}_2$~\cite{Breitenlohner:1982jf,Breitenlohner:1982bm}, exponentially growing unstable modes appear. This BF-bound violation is discussed in the context of the holographic superconductor~\cite{Hartnoll:2008kx}. While one may think that physical quantities do not blow
up without the violation of the BF bound, if the effective mass squared is in the range
\begin{equation}
\label{massrange2}
m^2_{{\rm BF},2}\le \frac{\bar{\mu}^2}{\lambda_0}<0,
\end{equation}
$\partial_\rho\phi_\ell|_{\rho=0}$ is divergent at the late time. Our analysis shows that the blowup of the first-order derivative implies that of the component of the energy-momentum tensor measured by the parallelly propagated null geodesic frame along extremal black hole horizons. 

Our analysis on the late-time tails is based on the near-horizon-geometry approximation of the field equation discussed in Sec.~\ref{subsec:MKGfiedsinNHG}. If we take into account the effect of subleading terms, we need to discuss the late-time behavior of the fields from the initial value problem. From this point of view, the late-time power-law tails are discussed for massless scalar fields near the horizon of the four-dimensional extremal Reissner-Nordstr\"om black holes in~\cite{Lucietti:2012xr,Bhattacharjee:2018pqb,Blaksley:2007ak}, and their results are consistent with the discussion from the near-horizon-geometry approximation. It is interesting to extend the analysis to our general setup,\footnote{
In the case of the massless scalar fields near the horizon of the four-dimensional extremal Reissner-Nordstr\"om black holes, the behaviors of the fields can be mapped into those near the infinity in the asymptotically flat spacetime using the discrete conformal isometry of the metric~\cite{Lucietti:2012xr,Bizon:2012we,Ori:2013iua,Sela:2015vua,Bhattacharjee:2018pqb,Couch:1984}.
Although there does not exist the global conformal isometry in our generic setup, 
the behaviors of the fields near the horizon still can be mapped into those near the infinity in the asymptotically flat spacetime using the conformal transformation, which maps the near horizon region into the asymptotically flat region. Thus, the same technique as~\cite{Ori:2013iua,Sela:2015vua,Bhattacharjee:2018pqb,Leaver:1986gd} can be used in our setup.}  but we leave this for future work.

\begin{acknowledgments}
The authors would like to thank Tomohiro Harada, Takaaki Ishii, Shunichiro Kinoshita, Tsutomu Kobayashi, and Norihiro Tanahashi for useful comments and discussions. This work was supported by Rikkyo University Special Fund for Research (TK) and  MEXT Grant-in-Aid for Scientific Research on Innovative Areas 20H04746 (MK).
\end{acknowledgments}

\appendix

\section{Explicit forms of $Z_N(\rho)$}
\label{Appendix:A}
Defining a function
\begin{align}
Z_N (\rho)= \sum_{i = 0}^{N} C_{N, i}^Z \rho^i,
\end{align}
we can choose the coefficients $C_{N, i}^Z$ such that
\begin{equation}
\begin{split}
\left.\partial_\rho^{N}\left[Z_N\left\{\partial_\rho\left(\rho^2\left(\lambda_0 +\delta\lambda\right)\left(\gamma_0+\delta\gamma\right)\partial_\rho\phi_\ell\right)-N(N+1)\lambda_0 \gamma_0 \phi_\ell-\mu^2\delta\gamma\phi_\ell\right\}\right]\right|_{\rho=0}=0,
\label{eq:ZeqAppendix}
\end{split}
\end{equation}
for arbitrary $\phi_\ell$.\footnote{We note that $\phi_\ell$ in this section is not necessarily a solution of the Klein-Gordon equation.}
 Expanding $\left(\lambda_0 +\delta\lambda\right)\left(\gamma_0+\delta\gamma\right)$ as
\begin{align}
\left(\lambda_0 +\delta\lambda\right)\left(\gamma_0+\delta\gamma\right) = 
 \sum_{i = 0}^{\infty} C_i^{\lambda\gamma} \rho^i
\end{align}
the coefficients $C_{N, i}^Z$ can be explicitly written in terms of $C_j^{\lambda\gamma}$ with $j \le i$.
Note that $C_0^{\lambda\gamma}= \lambda_0 \gamma_0$ and we can set $C^Z_{N,0} =1$.
We also expand $\phi_\ell$ as the Taylor series around $\rho = 0$
\begin{align}
\phi_\ell = \sum_{s = 0}^\infty \frac{1}{s!} \left.\partial_\rho^{s} \phi \right|_{\rho = 0} \rho^{s}.
\end{align}
After some calculation, we obtain
\begin{equation}
\begin{split}
& Z_N\left\{\partial_\rho\left[\rho^2\left(\lambda_0 +\delta\lambda\right)\left(\gamma_0+\delta\gamma\right)\partial_\rho\phi_\ell\right]-N(N+1)\lambda_0 \gamma_0\phi_\ell-\mu^2\delta\gamma\phi_\ell
\right\}\\=&
\sum_{i,j,s} \frac{s (1+s+i)}{s!}C_i^{\lambda\gamma} C^Z_{N, j} \left.\partial_\rho^{s} \phi \right|_{\rho = 0} \rho^{i+j+s}-\lambda_0 \gamma_0\sum_{j,s} \frac{N(N+1)}{s!} C^Z_{N, j} \left.\partial_\rho^{s} \phi \right|_{\rho = 0} \rho^{j+s}\\
&-\sum_{j,s,i} \frac{\mu^2}{s!} C^Z_{N, j} \gamma_i\left.\partial_\rho^{s} \phi \right|_{\rho = 0} \rho^{j+s+i},
\end{split}
\end{equation}
where $\gamma_i$ is defined in Eq.~\eqref{labellambdagamma}. Acting $\partial_\rho^N$ on this equation and evaluating it at $\rho = 0$, 
we obtain coefficients before $\rho^{N}$.
Then, Eq.~\eqref{eq:ZeqAppendix} becomes
\begin{equation}
\begin{split}
&\sum_{i=0}^N \sum_{j=0}^{N-i} 
\frac{(N-i-j) (1+N - j)}{(N-i-j)!} C_i^{\lambda\gamma} C^Z_{N, j} \left.\partial_\rho^{N-i-j} \phi \right|_{\rho = 0}-\lambda_0 \gamma_0\sum_{j=0}^N \frac{N(N+1)}{(N-j)!} C^Z_{N, j} \left.\partial_\rho^{N-j} \phi \right|_{\rho = 0}\\
&-\sum_{i=1}^N \sum_{j=0}^{N-i} \frac{\mu^2}{(N-j-i)!} C^Z_{N, j} \gamma_i\left.\partial_\rho^{N-j-i} \phi \right|_{\rho = 0}=0.
\end{split}
\end{equation}
Because this holds for arbitrary 
$\left.\partial_\rho^{j} \phi \right|_{\rho = 0}$ with $j = 0,1,\cdots, N$, we have
\begin{equation}
\begin{split}
\lambda_0 \gamma_0N(N+1) C^Z_{N, j}=&\sum_{i = 0}^j(N-j)(N+1+i-j)C_i^{\lambda\gamma} C^Z_{N, j-i}-\mu^2\sum_{i = 1}^jC^Z_{N, j-i}\gamma_i\\
=&\lambda_0 \gamma_0(N-j)(N+1-j) C^Z_{N, j}+\sum_{i = 1}^j (N-j)(N+1+i-j)C_i^{\lambda\gamma} C^Z_{N, j-i}\\
&-\mu^2\sum_{i = 1}^jC^Z_{N, j-i}\gamma_i.
\end{split}
\end{equation}
We here have used $C_0^{\lambda\gamma}=\lambda_0\gamma_0$ in the second line of the right-hand side. We thus obtain
\begin{equation}
\label{Zellrecursion}
C^Z_{N, i} =\frac{1}{\lambda_0 \gamma_0i(2N-i+1)} \left(\sum_{j = 1}^i (N-i)(N+1+j-i)C_j^{\lambda\gamma} C^Z_{N, i-j}-\mu^2\sum_{j = 1}^iC^Z_{N, i-j}\gamma_j\right).
\end{equation}
This shows that the coefficients $C^Z_{N,i}$ are expressed by $C^Z_{N,0} = 1$, the Taylor expansion coefficients of the metric, and the mass squared~$\mu^2$. For example, we find
\begin{equation}
C^Z_{1, 1}=-\frac{\mu^2\gamma_1}{2\lambda_0 \gamma_0},
\end{equation}
for $N=1$ and
\begin{equation}
C^Z_{2, 1}=\frac{3C_1^{\lambda\gamma}-\mu^2\gamma_1}{4\lambda_0 \gamma_0},~~C^Z_{2, 2}=-\mu^2\frac{3C_1^{\lambda\gamma}\gamma_1-\mu^2\gamma_1^2+4\lambda_0 \gamma_0\gamma_2}{24\left(\lambda_0 \gamma_0\right)^2},
\end{equation}
for $N=2$.

\section{Late-time power-law tails in terms of the near-horizon geometry}
\label{appendix:MKGfiedsinAdS2}
In this appendix, we discuss late-time behaviors of the massive Klein-Gordon field satisfying Eq.~\eqref{eom2} in terms of the near-horizon geometry. As shown in Sec.~\ref{subsec:MKGfiedsinNHG}, the late-time behavior of Eq.~\eqref{eom2} is described by the massive Klein-Gordon equation on ${\rm AdS}_2$,
\begin{equation}
\label{MKGAdS2inC}
\left[\square_{{\rm AdS}_2}-\bar{\mu}^2\right]\phi(v,\rho)=0,
\end{equation}
where $\bar{\mu}^2=\mu^2+\ell(\ell+n-3)/\gamma_0$. This is equivalent to Eq.~\eqref{eomtphi0}. We analyze Eq.~\eqref{MKGAdS2inC} by imposing normalizable boundary conditions and derive Eqs.~\eqref{latetimeouterthehorizon} and~\eqref{latetimeouterthehorizon2}. For simplicity, we assume the analyticity of $\phi$ at the horizon $\rho=0$.

\subsection{Specific mass squared case: $\bar{\mu}^2=\lambda_0 N(N+1)$}
We first discuss the specific mass squared case~$\bar{\mu}^2=\lambda_0 N(N+1)$ for a nonnegative integer~$N$, in which the Aretakis constants in ${\rm AdS}_2$ exist~\cite{Lucietti:2012xr,Cardoso:2017qmj,Katagiri:2021scx}. We expand $\phi$ in terms of $1/v$ as 
\begin{equation}
\label{rhoexpansion}
\phi=\sum_{j=0}^{\infty}\left(\frac{1}{v}\right)^j u_j(v\rho).
\end{equation}
where $u_j$ are functions of $v\rho$. Substituting this into Eq.~\eqref{MKGAdS2inC}, we obtain an equation for each $u_j$,
\begin{equation}
\label{equationforuj}
v\rho(2+\lambda_0v\rho)u_j''+2(1-j+\lambda_0v\rho)u_j'-\lambda_0 N(N+1)u_j=0,
\end{equation}
where the prime denotes the derivative with respect to $v\rho$. 

We further expand $u_j(v\rho)$ in Eq.~\eqref{equationforuj} as
\begin{equation}
\label{uexpansion}
u_j=\left(\frac{1}{2+\lambda_0 v\rho}\right)^j\sum_{k=0}^{\infty}c_{j,k}\left(\lambda_0 v\rho\right)^k,
\end{equation}
where $c_{j,k}$ are constants. Equation~\eqref{equationforuj} yields a recursion relation for $c_{j,k}$,
\begin{equation}
\label{recursionck}
\left(k-j+N+1\right)\left(k-j-N\right)c_{j,k}=-2\left(k+1\right)\left(k-j+1\right)c_{j,k+1}.
\end{equation}
For $j=0$, the left-hand side vanishes when $k=N$. Hence, we have a sequence of $c_{0,k}$ for $0\le k\le N$, while $c_{0,k}=0$ for $N+1\le k$. However, this sequence forms a finite polynomial in $v\rho$, and the field is divergent~$\phi\sim\rho^N$ at the AdS boundary $\rho=\infty$. Since we are interested in the solution satisfying normalizable boundary conditions, we consider $c_{0,k}=0$ for all $k$. 

For $1\le j\le N$, the left- and right-hand sides of Eq.~\eqref{recursionck} vanish when $k=j+N$ and $k=j-1$, respectively.  The former and latter imply $c_{j,k}=0$ for $j+N+1\le k$ and for $0\le k\le j-1$, respectively. Therefore, we have a single sequence of $c_{j,k}$ for $j\le k\le j+N$, which corresponds to non-normalizable solutions~$\phi\sim \rho^N$ at the AdS boundary $\rho=\infty$. We thus impose $c_{j,k}=0$ for $j\le k\le j+N$.\footnote{For $0\le j\le N$,  general solutions of Eq.~\eqref{equationforuj} are given  by
\begin{equation}
u_j=\left(\frac{v\rho}{2+\lambda_0 v\rho}\right)^{j/2}\left(c_P P_N^j(1+\lambda_0v\rho)+c_Q Q_N^j(1+\lambda_0v\rho)\right),
\end{equation}
where $c_P$, $c_Q$ are constants, and $P_N^j$, $Q_N^j$ are respectively the Legendre functions of the first kind and second kind. The sequence of $c_{j,k}$ for $j\le k\le j+N$ corresponds to the solution of $P_N^j$. However, the solutions of $Q_N^j$ are not analytic functions of $\rho$ at $\rho=0$ and therefore cannot be expressed by Eq.~\eqref{uexpansion}. For the solutions of $Q_N^j$, we have confirmed that $\partial_\rho^j u_j$ is divergent at $\rho=0$ for any $v$. In particular, for $j=0$, the value of the field~$u_0$ is divergent at the horizon. In this paper, we focus on the field~$\phi$ described by the analytic function at $\rho=0$. }

For $N+1\le j$, the left-hand side of Eq.~\eqref{recursionck} vanishes when $k=j-N-1$ and $k= j+N$, while the right-hand side does when $k=j-1$. The vanishing of the left- and right-hand sides at $k=j-N-1$ and $k=j-1$ imply that $c_{j,k}=0$  for $j-N\le k\le j-1$ when $1\le N$. Note that $c_{j,j-1}$ and  $c_{j,j}$ can be left arbitrary when $N=0$ because both sides of Eq.~\eqref{recursionck} vanish at $k=j-1$. The vanishing of the left-hand side at $k=j+N$  means that $c_{j,k}=0$ for $j+N+1\le k$. Therefore, we have two sequences for $0\le k\le j-N-1$ and $j\le k\le j+N$. The former sequence yields the solution that satisfies the normalizable condition, while the latter does not. We thus impose $c_{j,k}=0$ for $N+1\le j\le k$. 

To summarize, we have the following solution satisfying the normalizable boundary condition\footnote{\label{footnote:generalsols}On a null hypersurface~$v=v_0$, choosing $c_{N+1+s,k}$ appropriately, $\phi$ in Eq.~\eqref{phigeneralsolution} can be any  function of $\rho$ on $v=v_0$. This implies that $\phi$ in Eq.~\eqref{phigeneralsolution} is a general solution which satisfies the normalizable boundary condition.}
\begin{equation}
\label{phigeneralsolution}
\phi=\sum_{s=0}^{\infty}\sum_{k=0}^{s}c_{N+1+s,k}\left(\frac{1}{v}\right)^{N+1+s} \left(\frac{1}{2+\lambda_0 v\rho}\right)^{N+1+s}\left(\lambda_0 v\rho\right)^k.
\end{equation}
In particular, the leading contribution at the late time $v\to\infty$ is described by the mode with $k=s=0$,
\begin{equation}
\label{latetimeouterthehorizonAdSN}
\phi\simeq\frac{N!}{\lambda_0^{N+1}(2N+1)!}\mathcal{H}_N^{{\rm AdS}_2}\left(-\frac{v}{2}-\frac{\lambda_0\rho v^2}{4}\right)^{-N-1}.
\end{equation}
Here, we have chosen $c_{N+1,0}$ as
\begin{equation}
c_{N+1,0}=\frac{N!}{(2N+1)!}\left(-\frac{4}{\lambda_0}\right)^{N+1}\mathcal{H}_N^{{\rm AdS}_2},
\end{equation}
so that $\partial_\rho^N\phi|_{\rho=0}=\mathcal{H}_N^{{\rm AdS}_2}$, which is the Aretakis constant in ${\rm AdS}_2$. 
This is consistent with the result  in~\cite{Lucietti:2012xr}.

\subsection{General mass squared cases with $\bar{\mu}^2\ge -\lambda_0/4$}
We parametrize the mass squared as $\bar{\mu}^2=\lambda_0\Delta(\Delta-1)$ with $\Delta \ge 1/2$ so that $\bar{\mu}^2$ is greater than or equal to $-\lambda_0/4$, which is the Breitenlohner-Freedman bound in ${\rm AdS}_2$. The case where $\Delta$ is an integer is included in the previous subsection with $N = \Delta - 1$. In this subsection, we assume that $\Delta$ is not an integer. We expand $\phi$ as
\begin{equation}
\label{rhoexpansionDelta}
\phi=\sum_{j=0}^{\infty}\left(\frac{1}{v}\right)^{j+b}u_{j+b}(v\rho),
\end{equation}
where $u_{j+b}$ are functions of $v\rho$ and $b=\Delta-\lfloor \Delta\rfloor$, where $\lfloor \Delta\rfloor$ denotes the integer part of $\Delta$.\footnote{If we set $b = 0$ for non-integer $\Delta$ case, the solution in Eq.~\eqref{rhoexpansionDelta} can describe only non-normalizable modes.} 
For later convenience, we introduce an integer~$\mathcal{N}=\Delta-b-1=\lfloor\Delta\rfloor-1$. Note that $\mathcal{N} \ge -1$ because $ \Delta\ge 1/2$. When $1/2 \le \Delta <1$, the mass squared $\bar{\mu}^2$ and $b$ satisfy $-\lambda_0/4 \le \bar{\mu} ^2 <0$, $1/2 \le b <1$, and ${\cal N}$ becomes $-1$.

Substituting Eq.~\eqref{rhoexpansionDelta} into Eq.~\eqref{MKGAdS2inC}, we obtain an equation for $u_{j+b}$, 
\begin{equation}
\label{equationforujb}
v\rho(2+\lambda_0v\rho)u_{j+b}''+2(1-j-b+\lambda_0v\rho)u_{j+b}'-\lambda_0 (\mathcal{N}+1+b)(\mathcal{N}+b)u_{j+b}=0.
\end{equation}
The general solutions are
\begin{equation}
\begin{split}
\label{generalujb}
u_{j+b}=&c_{F1} \!~_2F_1\left(\mathcal{N}+1+b,-\mathcal{N}-b;1-j-b;-{\lambda_0v\rho}/{2}\right)\\
&+c_{F2} \left(-\frac{\lambda_0 v\rho}{2}\right)^{j+b}\!~_2F_1\left(\mathcal{N}+1+j+2b,-\mathcal{N}+j;1+j+b;-{\lambda_0v\rho}/{2}\right),
\end{split}
\end{equation}
where $c_{F1}$, $c_{F2}$ are constants and $\!~_2F_1$ is the Gaussian hypergeometric function. The asymptotic behaviors near the AdS boundary~$\rho=\infty$ take the form~\cite{Abramowitz:1972}
\begin{equation}
\begin{split}
u_{j+b}=&\frac{\Gamma\left(2\mathcal{N}+1+b\right)}{\Gamma\left(\mathcal{N}+1+b\right)}\left[\frac{c_{F1}\Gamma\left(1-j-b\right)}{\Gamma\left(\mathcal{N}+1-j\right)}+\frac{c_{F2}(-1)^{j+b}\Gamma\left(1+j+b\right)}{\Gamma\left(\mathcal{N}+1+j+2b\right)}\right]\left(\frac{\lambda_0 v\rho}{2}\right)^{\mathcal{N}+b}+\mathcal{O}\left(\rho^{-(\mathcal{N}+1+b)}\right).
\end{split}
\end{equation}
This implies that the field~$\phi$ satisfies the normalizable boundary conditions in two cases: (i)~$j\ge \mathcal{N}+1$ and $c_{F2}=0$, and (ii)~$j< \mathcal{N}+1$ and
\begin{equation}
c_{F_2}=(-1)^{j+b+1}\frac{\Gamma\left(\mathcal{N}+1+j+2b\right)\Gamma\left(1-j-b\right)}{\Gamma\left(\mathcal{N}+1-j\right)\Gamma\left(1+j+b\right)}c_{F1}.
\end{equation}
However, in the case~(ii), the term of $c_{F2}$  is not analytic at $\rho=0$ due to the presence of the factor with the fractional power, $\rho^{j+b}$, of which the derivatives of $u_{j+b}$ with respect to $\rho$ is divergent at $\rho=0$ for any $v$. Since we are considering the analytic solution, we discard the case~(ii).

To obtain further perspective of the case~(i), we expand $u_{j+b}$, where $j\ge \mathcal{N}+1$, as
\begin{equation}
\label{uexpansionDelta}
u_{j+b}=\left(\frac{1}{2+\lambda_0 v \rho}\right)^{j+b} \sum_{k=0}^{\infty} c_{j,k}(\lambda_0v\rho)^k.
\end{equation}
Substituting this into Eq.~\eqref{equationforujb}, we obtain a recursion relation for $c_{j,k}$,
\begin{equation}
\label{recursionck3}
\left(k-j+\mathcal{N}+1\right)\left(k-j-\mathcal{N}-2b\right)c_{j,k}=-2\left(k+1\right)\left(k-j-b+1\right)c_{j,k+1}.
\end{equation}
We notice that the right-hand side of Eq.~\eqref{recursionck3} does never vanish as long as $c_{j,k+1}\neq 0$. Because of $j\ge \mathcal{N}+1$ in the current case, the left-hand side vanishes only when $k=j-\mathcal{N}-1$ if $b\neq 1/2$ and also when $k=j+\mathcal{N}+1$ if $b=1/2$. In any case, there is a single finite sequence of $c_{j,k}$ for $0\le k\le \mathcal{N}$. This forms a finite polynomial in $\lambda_0v\rho$ and gives rise to $\phi$ satisfying the normalizable boundary condition at the AdS boundary $\rho=\infty$. This solution should be proportional to the first term in Eq.~\eqref{generalujb} because the same boundary conditions are satisfied.
 
To summarize, we have the general solutions satisfying the normalizable condition,\footnote{For the same reason as mentioned in  footnote~\ref{footnote:generalsols}, Eq.~\eqref{generalphiingeneralmass} is a general solution satisfying the normalizable boundary condition.}
\begin{equation}
\label{generalphiingeneralmass}
\phi=\sum_{s=0}^{\infty}\sum_{k=0}^{s} c_{\mathcal{N}+1+b+s,k}\left(\frac{1}{v}\right)^{\mathcal{N}+1+b+s}\left(\frac{1}{2+\lambda_0 v \rho}\right)^{\mathcal{N}+1+b+s} (\lambda_0v\rho)^k.
\end{equation}
Using this form of solutions rather than the form in the hypergeometric function in Eq.~\eqref{generalujb}, we can estimate the leading contribution at the late time $v \to \infty$ is described by the $s = k = 0$ mode in Eq.~\eqref{generalphiingeneralmass},
\begin{equation}
\label{generalsolphic}
\phi\simeq c_{\mathcal{N}+1+b,0}\left(\frac{1}{2v+\lambda_0 v^2 \rho}\right)^{\mathcal{N}+1+b}.
\end{equation}
Note that the power~$\mathcal{N}+1+b=\Delta\ge 1/2$. Defining a constant $H_\Delta^{{\rm AdS}_2}$ by
\begin{equation}
c_{\mathcal{N}+1+b,0}=\frac{\Gamma(\mathcal{N}+1+b)}{\Gamma(2\mathcal{N}+2+2b)}\left(-\frac{4}{\lambda_0}\right)^{\mathcal{N}+1+b} H_\Delta^{{\rm AdS}_2},
\end{equation}
Eq.~\eqref{generalsolphic} is rewritten as
\begin{equation}
\label{latetimeouterthehorizon2AdSDelta}
\phi\simeq\frac{\Gamma(\mathcal{N}+1+b)}{\lambda_0^{\mathcal{N}+1+b}\Gamma(2\mathcal{N}+2+2b)}H_\Delta^{{\rm AdS}_2}\left(-\frac{v}{2}-\frac{\lambda_0\rho v^2}{4}\right)^{-\mathcal{N}-1-b}.
\end{equation}
This equation corresponds to Eq.~\eqref{latetimeouterthehorizonAdSN} and coincides with it in the limit of $b\to 0$. This is consistent with~\cite{Lucietti:2012xr}.

\section{Derivation of Eq.~\eqref{relationP}}
\label{Appendix:P}
Acting $\partial_\rho^N$ on Eq.~\eqref{deltaN}, we obtain
\begin{equation}
\begin{split}
\label{dNdelta}
\gamma_0\partial_{\rho}^{N}\delta[\phi_\ell]=&-\left.\partial_v\partial_\rho^{N}\left[2\delta\gamma\partial_\rho\phi_\ell+\left(\partial_\rho\delta\gamma\right)\phi_\ell\right]\right|_{\rho=0}\\
&-\left.\partial_\rho^{N+1}\left[\rho^2(\gamma_0\delta\lambda+\lambda_0 \delta\gamma+\delta\lambda\delta\gamma)\partial_\rho\phi_\ell\right]\right|_{\rho=0}+\left.\mu^2\partial_\rho^N\left(\delta\gamma\phi_\ell\right)\right|_{\rho=0}\\
&+\mathcal{O}\left(\rho\right),
\end{split}
\end{equation}
near $\rho=0$. Using the Leibniz rule for some functions $\mathcal{F}_1$ and $\mathcal{F}_2$, 
\begin{equation}
\partial_\rho^j\left(\mathcal{F}_1\mathcal{F}_2\right)=\sum_{i=0}^j\frac{j!}{i!(j-i)!}\left(\partial_\rho^i\mathcal{F}_1\right)\left(\partial_\rho^{j-i}\mathcal{F}_2\right),
\end{equation}
Eq.~\eqref{dNdelta} is rewritten as
\begin{equation}
\begin{split}
\label{dNdelta2}
\gamma_0\partial_{\rho}^{N}\delta[\phi_\ell]=&-\left.\partial_v\partial_\rho^{N}\left[2\delta\gamma\partial_\rho\phi_\ell+\left(\partial_\rho\delta\gamma\right)\phi_\ell\right]\right|_{\rho=0}\\
&-\sum_{i=0}^{N+1}\frac{(N+1)!}{i!(N+1-i)!}\left.\partial_\rho^{i}\left[\rho^2(\gamma_0\delta\lambda+\lambda_0 \delta\gamma+\delta\lambda\delta\gamma)\right]\partial_\rho^{N+2-i}\phi_\ell\right|_{\rho=0}\\
&+\mu^2\sum_{i=0}^{N}\frac{N!}{(N-i)!}\gamma_i\partial_\rho^{N-i}\left.\phi_\ell\right|_{\rho=0}\\
&+\mathcal{O}\left(\rho\right).
\end{split}
\end{equation}
Here, from the $i$th-order derivatives of Eq.~\eqref{eom2} with respect to $\rho$, we notice that $\partial_\rho^{i}\phi_\ell|_{\rho=0}$ has the following relation:
\begin{equation}
\label{Gjrelation}
\partial_v\mathcal{G}_{i}=\left.\partial^{i}_\rho\phi_\ell\right|_{\rho=0}.
\end{equation}
We have here defined
\begin{equation}
\left.\mathcal{G}_{N-j}(v):=\frac{2}{\lambda_0 \gamma_0j(2N-j+1)}\partial_{\rho}^{N-j}\left[\left(\gamma_0+\delta\gamma\right)^{1/2}\partial_\rho\left\{\left(\gamma_0+\delta\gamma\right)^{1/2}\phi_\ell\right\}\right]\right|_{\rho=0},
\end{equation}
which corresponds to $\mathcal{G}_{j}(v)$ in Eq.~\eqref{Gj}. With the relation~\eqref{Gjrelation}, Eq.~\eqref{dNdelta2} can be rewritten as 
\begin{equation}
\label{relationPinapp}
\partial_{\rho}^{N}\delta[\phi_\ell]=\partial_vP_N+\mathcal{O}\left(\rho\right),
\end{equation}
where
\begin{equation}
\begin{split}
\label{functionPinapp}
\gamma_0P_N(v)=&-\left.\partial_\rho^{N}\left[2\delta\gamma\partial_\rho\phi_\ell+\left(\partial_\rho\delta\gamma\right)\phi_\ell\right]\right|_{\rho=0}\\
&-\sum_{i=0}^{N+1}\frac{(N+1)!}{i!(N+1-i)!}\left.\partial_\rho^i\left[\rho^2\left(\gamma_0\delta\lambda+\lambda_0 \delta\gamma+\delta\lambda\delta\gamma\right)\right]\right|_{\rho=0}\mathcal{G}_{N+2-i}(v)\\
&+\mu^2\sum_{i=0}^{N}\frac{N!}{(N-i)!}\gamma_i\mathcal{G}_{N-i}(v).
\end{split}
\end{equation}

\section{Proof of Eq.~\eqref{DDDdelta}}
\label{Appendix:B}
In this Appendix, we introduce symbolic forms,
\begin{equation}
\begin{split}
\delta_{N-2}=&D_{N-2}\delta\left[\phi_\ell\right],\\
\delta_{N-3}=&D_{N-3}\delta_{N-2},\\
&~~\vdots\\
\delta_{k}=&D_{k}\delta_{k+1},\\
&~~\vdots\\
\delta_{0}=&D_{0}\delta_{1},\\
\delta_{-1}=&D_{-1}\delta_{0}.
\end{split}
\end{equation}
Here, we have set $\delta_{N-1}=\delta\left[\phi_\ell\right]$, where $\delta\left[\phi_\ell\right]$ is defined in Eq.~\eqref{deltaN}. In this notation, the left-hand side of Eq.~\eqref{DDDdelta} is $\delta_{-1}$.

The function~$\delta_{-1}$ is explicitly calculated to
\begin{equation}
\label{k0}
\delta_{-1}=\partial_v\left(v^2\delta_0\right)+\frac{2}{\lambda_0 }\partial_\rho\delta_0+\mathcal{O}\left(\rho\right).
\end{equation}
We shall prove the following relation with mathematical induction:
\begin{equation}
\label{deltak}
\delta_{-1}=\partial_vW_i+\left(\frac{2}{\lambda_0 }\right)^{i+1}\partial_\rho^{i+1}\delta_i+\mathcal{O}\left(\rho\right),
\end{equation}
where $W_i=W_i(v,\rho)$ is some function. For $i=0$, it is clear that the relation~\eqref{deltak} holds because of Eq.~\eqref{k0} with $W_0=v^2\delta_0$. We now assume that the relation~\eqref{deltak} holds for $i=j$. We calculate
\begin{equation}
\begin{split}
\label{deltam}
\partial_\rho^{j+1}\delta_{j}=&\frac{2}{\lambda_0 }\partial_\rho^{j+2}\delta_{j+1}+v^2\partial_v\partial_\rho^{j+1}\delta_{j+1}-2jv\partial_\rho^{j+1}\delta_{j+1} \\
&+2v\partial_\rho^{j+1}\left(\rho\partial_\rho\delta_{j+1}\right)+\lambda_0 v^2\partial_\rho^{j+1}\left(\rho^2\partial_\rho\delta_{j+1}\right)-\lambda_0 jv^2\partial_\rho^{j+1}\left(\rho\delta_{j+1}\right).
\end{split}
\end{equation}
This is rewritten as
\begin{equation}
\begin{split}
\partial_\rho^{j+1}\delta_{j}=&\partial_v\left(v^2\partial_\rho^{j+1}\delta_{j+1}\right)+\frac{2}{\lambda_0 }\partial_\rho^{j+2}\delta_{j+1}+\mathcal{O}\left(\rho\right),
\end{split}
\end{equation}
with the aid of relations
\begin{equation}
\begin{split}
\label{formulaF}
\partial_\rho^{j+1}\left(\rho\mathcal{F}\right)=&\left(j+1\right)\partial_\rho^{j}\mathcal{F}+\mathcal{O}\left(\rho\right)\\
\partial_\rho^{j+1}\left(\rho^2\mathcal{F}\right)=&j\left(j+1\right)\partial_\rho^{j-1}\mathcal{F}+\mathcal{O}\left(\rho\right),
\end{split}
\end{equation}
where $\mathcal{F}=\mathcal{F}(v,\rho)$ is an analytic function. Then, it follows that
\begin{equation}
\begin{split}
\label{deltamplus1}
\delta_{-1}=&\partial_v W_j+2^{j+1}\partial_\rho^{j+1}\delta_j+\mathcal{O}\left(\rho\right)\\
=&\partial_vW_{j+1}+\left(\frac{2}{\lambda_0 }\right)^{j+2}\partial_\rho^{j+2}\delta_{j+1}+\mathcal{O}\left(\rho\right),
\end{split}
\end{equation}
where we have defined 
\begin{equation}
\label{Wm}
W_{j+1}=W_j+\left(\frac{2}{\lambda_0 }\right)^{j+1}v^2\partial_\rho^{j+1}\delta_{j+1}.
\end{equation}
It can be seen that the relation~\eqref{deltak} also holds for $i=j+1$. Thus, we have shown the relation~\eqref{deltak}. 

When $i=N-1$ in the relation~\eqref{deltak}, we have 
\begin{equation}
\label{deltaminus1}
\delta_{-1}=\partial_vW_{N-1}+\left(\frac{2}{\lambda_0 }\right)^{N}\partial_\rho^{N}\delta\left[\phi_\ell\right]+\mathcal{O}\left(\rho\right),
\end{equation}
where we have used $\delta_{N-1}=\delta\left[\phi_\ell\right]$. Using the recursion~\eqref{Wm}, we can explicitly obtain
\begin{equation}
\begin{split}
W_{N-1}=&W_{N-2}+\left(\frac{2}{\lambda_0 }\right)^{N-1}v^2\partial_\rho^{N-1}\delta_{N-1}\\
=&W_{N-3}+\left(\frac{2}{\lambda_0 }\right)^{N-2}v^2\partial_\rho^{N-2}\delta_{N-2}+\left(\frac{2}{\lambda_0 }\right)^{N-1}v^2\partial_\rho^{N-1}\delta_{N-1}\\
=&v^2\delta_0+\left(\frac{2}{\lambda_0 }\right)v^2\partial_\rho\delta_1+\left(\frac{2}{\lambda_0 }\right)^2v^2\partial_\rho^{2}\delta_2+\cdots+\left(\frac{2}{\lambda_0 }\right)^{N-1}v^2\partial_\rho^{N-1}\delta_{N-1},
\end{split}
\end{equation}
where we have used $W_0=v^2\delta_0$. Furthermore, using $P_N$ in Eq.~\eqref{functionP}, the relation~\eqref{deltaminus1} is rewritten as 
\begin{equation}
\begin{split}
\delta_{-1}=&\partial_v\left[v^2\delta_0+\cdots+\left(\frac{2}{\lambda_0 }\right)^{N-2}v^2\partial_\rho^{N-2}\delta_{N-2}+\left(\frac{2}{\lambda_0 }\right)^{N-1}v^2\partial_\rho^{N-1}\delta\left[\phi_\ell\right]+\left(\frac{2}{\lambda_0 }\right)^{N}P_N\right]+\mathcal{O}\left(\rho\right).
\end{split}
\end{equation}

\section{Explicit calculations of the Aretakis constants}
\label{appendix:Examples}
\subsection{Aretakis constant for $N=1$}
This corresponds to the case $\mu^2=2\lambda_0 -\ell(\ell+n-3)/\gamma_0$. The quantity~$\mathcal{Q}_1$ in Eq.~\eqref{QN} is explicitly calculated to
\begin{equation}
\begin{split}
\label{Q1}
\frac{\gamma_0}{2}\mathcal{Q}_1=&\left.\partial_\rho^{2}\phi_\ell\right|_{\rho=0}+\frac{3\lambda_0 -\mu^2}{2\lambda_0 \gamma_0}\gamma_1\left.\partial_\rho\phi_\ell\right|_{\rho=0}+\frac{4\lambda_0 \gamma_0\gamma_2-\mu^2\gamma_1^2}{4\lambda_0 \gamma_0^2}\left.\phi_\ell\right|_{\rho=0}.
\end{split}
\end{equation}
We have here used Eq.~\eqref{eom2} with $N=1$ at $\rho=0$. The right-hand side of this equation coincides with that of Eq.~\eqref{garec} with $N=1$. Thus, $(\gamma_0/2)\mathcal{Q}_1$ is the Aretakis constant $\mathcal{H}_{1}$. 

\subsection{Aretakis constant for $N=2$}
This corresponds to the case $\mu^2=6\lambda_0 -\ell(\ell+n-3)/\gamma_0$. The quantity~$\mathcal{Q}_2$ in Eq.~\eqref{QN} is explicitly calculated to
\begin{equation}
\begin{split}
\label{Q2}
&\left(\frac{\lambda_0 }{2}\right)^{2}\mathcal{Q}_2\\
=&\left.\partial_\rho^{3}\phi_\ell\right|_{\rho=0}+\left(\frac{8\lambda_0 -\mu^2}{2\lambda_0 \gamma_0}\gamma_1+\frac{3}{2\lambda_0 }\lambda_1\right)\left.\partial_\rho^{2}\phi_\ell\right|_{\rho=0}\\
&+\left(\frac{27\lambda_0 ^2\gamma_0-12\lambda_0 \mu^2+\mu^4}{12\lambda_0 ^2\gamma_0^2}\gamma_1^2+\frac{12\lambda_0 ^2-\mu^2}{3\lambda_0 \gamma_0}\gamma_2+\frac{9\lambda_0 -\mu^2}{4\lambda_0 ^2\gamma_0}\gamma_1\lambda_1\right)\left.\partial_\rho\phi_\ell\right|_{\rho=0}\\
&+\left[\frac{3}{\gamma_0}\gamma_3+\frac{9\lambda_0 -4\mu^2}{6\lambda_0 \gamma_0^2}\gamma_1\gamma_2\right.\\
&~~~~~+\left.\frac{3}{2\lambda_0 \gamma_0}\lambda_1\gamma_2-\frac{\left(3\lambda_0 -\mu^2\right)\gamma_1+3\gamma_0\lambda_1}{24\lambda_0 ^2\gamma_0^3}\mu^2\gamma_1^2\right]\left.\phi_\ell\right|_{\rho=0}.
\end{split}
\end{equation}
We have used the first-order derivative of Eq.~\eqref{eom2} with respect to $\rho$ and $v$ at $\rho=0$. The right-hand side of Eq.~\eqref{Q2} coincides with  that of Eq.~\eqref{garec} with $N=2$. Thus, $(\lambda_0 /2)^{2}\mathcal{Q}_2$ is nothing but the Aretakis constant $\mathcal{H}_{2}$.

\end{document}